\def \kev{~\rm{keV}}

\def \cm{~\rm{cm}}
\def \s{~\rm{s}}
\def \km{~\rm{km}}

\def \K{~\rm{K}}

\def \AU{~\rm{AU}}
\def \erg{~\rm{erg}}

\def \yr{~\rm{yr}}

\def \kpc{~\rm{kpc}}
\def \etc{$\eta$~Car}
\documentclass[12pt,preprint]{aastex}

\shorttitle{Eta Carinae Binarity} \shortauthors{Soker}
\slugcomment{Draft version of \today}

\begin{document}

\title{ACCRETION ONTO THE COMPANION OF ETA CARINAE DURING THE SPECTROSCOPIC
EVENT: II. X-RAY EMISSION CYCLE}

\author{Muhammad Akashi, Noam Soker, Ehud Behar \altaffilmark{}}
\affil{Department of Physics, Technion$-$Israel
Institute of Technology, Haifa 32000 Israel;
akashi@techunix.technion.ac.il; soker@physics.technion.ac.il;
behar@physics.technion.ac.il.}

\begin{abstract}

We calculate the X-ray luminosity and light curve for the stellar
binary system \etc\ for the entire orbital period of 5.54 years.
By using a new approach we find, as suggested before, that the
collision of the winds blown by the two stars can explain the
X-ray emission and temporal behavior. Most X-ray emission
{{{{{ in the $2-10 \kev$ band }}}}} results
from the shocked secondary stellar wind. The observed rise in
X-ray luminosity just before minimum is due to
{{{{{ increase in density and subsequent }}}}}
decrease in radiative cooling time of the shocked fast secondary wind.
Absorption, particularly of the soft X-rays from the primary wind, increases as
the system approaches periastron and the shocks are produced deep inside the
primary wind. However, absorption can not account for the drastic X-ray
minimum. The 70 day minimum is assumed to result
from the collapse of the collision region of the two winds onto
the secondary star. This process is assumed to shut down the
secondary wind, hence the main X-ray source. We show that this
assumption provides a phenomenological description of the
X-ray behavior around the minimum.

\end{abstract}

\keywords{binaries: close$-$circumstellar matter$-$stars:
individual: $\eta$ Carinae$-$stars: mass loss$-$stars: winds}

\section{INTRODUCTION}
\label{sec:intro}

The X-ray light curve of $\eta$ Carinae has a period of
5.54~years with a deep, non-zero minimum of $\sim 70$~days (Corcoran
2005) every period.
Some variability is observed during the minimum (Hamaguchi et
al. 2005).
The minimum is more or less coincidental with the fading of many
visible emission lines and a decline in the IR. It is widely
accepted that $\eta$ Car is a binary system (e.g. Damineli 1996;
Damineli et al.\ 1997, 2000; Ishibashi et al.\ 1999; Corcoran et
al.\ 2001a,b; 2004b; Pittard \& Corcoran 2002; Duncan \& White
2003;, Fernandez Lajus et al.\ 2003; Smith et al.\ 2004; Whitelock
et al.\ 2004; Steiner \& Damineli 2004; Verner et al.\ 2005;
{{{{Iping et al. 2005), }}}} and we refer to it as such. The more
massive companion of the $\eta$ Car binary system {{{ (the one
that caused the Great Eruption of $\sim 1840$; see description of
the eruption in Davidson \& Humphreys 1997) }}} will be referred
to here as the primary, while the companion, probably an O-type
star (Verner et al. 2005), will be referred to as the secondary.
The aforementioned periodic brightness minima are believed to
occur near periastron passage and are generally termed the
spectroscopic event (e.g., Damineli et al.\ 2000).

Corcoran et al.\ (2001a) and Pittard \& Corcoran (2002)
demonstrated that the collision of the winds blown by the two
stars can account for the X-ray light curve. However, they had
problems in accounting for the X-ray minimum. The X-ray
properties of $\eta$ Car are reviewed in section 2.1. In examining
the properties of the colliding winds of the two stars, Soker
(2005a, b) has suggested that for several weeks near periastron
passages the secondary accretes mass from the primary. {{{ This
suggestion of accretion is motivated by observations, which show
accretion from winds in many other types of binary systems, e.g.,
symbiotic systems (e.g., Skopal 2005), and from numerical
simulations showing the likelihood of accretion from a wind (e.g.,
Jahanara, et al. 2005), where in some cases an accretion disk is
formed  (Mitsumoto et al. 2005). }}} This accretion, Soker
suggested, shuts down the secondary's wind, hence the X-ray
emission, leading to the X-ray minimum. This scenario is
described in section 2.2. However, Soker did not explore the X-ray
light curve around the minimum. This is the subject of the present
paper. In \S 3 the X-ray emission of the colliding winds is
examined. The X-ray emission during the accretion phase is
examined in \S 4. Our discussion, where the model for the X-ray
minimum is summarized according to the results of \S 3 and 4, is
presented in \S 5.

\section{X-RAY MINIMUM}
\label{sec:modelX}
\subsection{Observed X-Ray Minimum}
\label{sec:minimum}
The main relevant properties of the X-ray emission near
the minimum are as follows.

{\it (1) Minimum.} The X-ray emission of \etc\ between 2 and
10 keV is normally at a level of 4$\times$10$^{34}$~erg~s$^{-1}$.
This emission is believed to result from the collision of the two
winds from the two stars (Corcoran et al.\ 2001a; Pittard \&
Corcoran 2002), as in similar massive binary systems (Usov 1992),
most notably the WR-O binary system WR~140 (Williams et al.\
1990). There is an {{{ almost }}} flat minimum in the
X-ray emission of \etc, lasting $\sim 70$~days (0.035 of the
cycle) (Ishibashi et al.\ 1999; Corcoran et al.\ 2001a; Corcoran
et al.\ 2004b; Corcoran 2005). The interacting massive binary
system WR 140 shows a similar X-ray light curve to that of
$\eta$~Car, but the minimum {{{ is very short and not as
flat.}}} Hence, it can be explained by absorption { of the
X-rays} by the dense wind of the WR star (Corcoran et al.\ 2004b;
Pollock et al.\ 2005).
Conversely, in $\eta$ Car the X-ray minimum is
{{{ almost }}} flat and appears to be a reduction in observed X-ray
emission measure, which is difficult to explain by absorption alone
(Ishibashi et al.\ 2003; Hamaguchi et al.\ 2005)

{\it (2) Behavior prior to minimum.} The X-ray intensity increases
prior to the start of the X-ray minimum and then drops sharply to
minimum (Corcoran 2005).
In this paper, phases are measured from periastron passage,
which may or may not be the same as the start of the X-ray
minimum (Martin et al.\ 2005).
The decline starts before the maximum of the
\ion{He}{2} $\lambda~4686$\AA\ line (Martin et al.\ 2005) and the
maximum of the IR (Whitelock et al.\ 2004). The X-ray decline
starts just after phase 0.98 (or $-0.02$); In the 2003 minimum,
there was an X-ray flare at phase $0.99$ (Corcoran 2005). The
minimum starts just before phase zero, and ends at phase $\sim
0.03$ (or $\sim 1.03$). The sharp decline in the IR and HeII
$\lambda~4686$\AA\ line starts at phase $\sim 0$.

{\it (3) No hardening during the minimum.} For $\sim 40$~days
into the minimum, the RXTE-observed $2-10 \kev$ X-ray
luminosity is $\sim 3 \times 10^{33} \erg \s^{-1}$ (for a distance
of 2.3~\kpc\ to $\eta$ Car), with large relative variations,
{{{ e.g., spikes (Corcoran 2005). However, the RXTE field of
view includes both the central source and the surrounding nebula.
The X-ray emission by the extended nebula is much softer than
emission from the stellar source (Hamaguchi et al. 2005),
hence RXTE finds a softer spectrum during
the minimum (Corcoran 2005), which is not intrinsic to the
central source. Hamaguchi et al. (2005) observed \etc\ 24 days
into the minimum (on 2003-07-22) with {\it XMM-Newton}.
Despite the huge decline, the X-ray spectrum does not become
harder. This shows that absorption cannot be the main reason for
the X-ray decline. After another several days the X-ray luminosity
starts to increase for a while (Corcoran 2005), and becomes harder
(Corcoran 2005; Hamaguchi et al. 2005). This again contradicts an
absorption effect in which a rise in flux (less absorption) would
be accompanied by softening.}}} We note that the softening in the
RXTE plot of Cororan (2005) starts a little after the decline
starts. {{{ If the hardness variation was only due to the
central source diminishing while the surroundings remain constant,
then the total radiation observed by RXTE should have become
softer at the same time the decline started. This shows that
although absorption is not the main factor during the
minimum, it may still play a role during the decline to, and the
rise out of, the  minimum.
}}} About 30~days into the  X-ray minimum, RXTE detected an
increase in X-ray flux to $\sim 6 \times 10^{33} \erg \s^{-1}$.
The spectrum becomes even a little harder (Corcoran 2005). This is
followed by a slow decline back to $\sim 4 \times 10^{33} \erg
\s^{-1}$ and by softening of the emission, before the steep rise
out of the minimum.

{\it (4) Flares.} {{{ Flares occur during most of the orbit.
Their effect in the RXTE $2-10 \kev$ band is a $\sim 10 \%$
increase in luminosity. This may indicate stochastic variations in
one or two of the wind parameters (velocity; density;
inhomogeniety), and/or instabilities in the wind interaction
region. The flare amplitudes increase somewhat with the X-ray
average luminosity after phase $\sim 0.6$ (Corcoran 2005). }}}
However, during the maximum period before decline, there are very
strong variations, with ``flares'' which increase the intensity by
up to $\sim 50 \%$ (Corcoran 2005). {{{ Such strong flares most
likely indicate much stronger instabilities in the winds
interaction process. Such instabilities may lead to the formation
of dense blobs in the post-shock primary wind region, which
eventually are favorable for accretion by the secondary (Soker
2005b).
}}}

{\it (5) Cycle-to-cycle variation.} {{{ Two minima were
observed in X-rays by RXTE. The X-ray luminosity at the maximum emission
just prior to the X-ray minimum in 2003.5 is larger that that in
1997.9 (Corcoran 2005). The qualitative behavior is similar in the
two minima, however, with similar time scales. }}}

\subsection{Proposed Model}
\label{sec:model}
In order for absorption to account for the deep X-ray minimum,
Corcoran et al.\ (2001a) were required to assume that the mass
loss rate from the primary in $\eta$~Car increases by a factor of
$\sim 20$ for 80 days following periastron. This model also
requires the secondary to be behind the primary during periastron
(Corcoran et al.\ 2004b). {{{ Ishibashi (2001) first showed
that a semi-major axis orientation perpendicular to our line of sight
provided the best fit to the asymmetry in the X-ray light curve
before and after the event, if the secondary is behind the primary
after the event. }}} We note that in a recent paper Smith et al.\
(2004) argue that the semi-major axis of the $\eta$ Car binary system
is indeed more or less perpendicular to the line of sight, but the
companion is not behind the primary near or after periastron,
{{{ but rather the secondary is behind the primary before
periastron passage. }}} Falceta-Goncalves et al.\ (2005) {{{
and Abraham et al. (2005) }}} build a different model, where the
secondary is in-front of the primary during periastron; this model
requires that we observe the $\eta$ Car system in the orbital
plane. However, from the structure of the Homunculus we believe the
orbital plane is tilted by {{{  $\sim 48 ^\circ$ from an
edge-on view {{{{{ (an inclination angle of $i=42^\circ$; }}}}}
Smith 2002). }}} It appears that the model of
Falceta-Goncalves et al.\ (2005) can not work for an inclined
system, although they do not give enough details for a conclusive
assessment.

In the presently proposed model, the collapse of the wind
collision region onto the secondary is the reason for the sharp
decline in the X-ray intensity. Figure \ref{forbit1} shows the
proposed orbit in the top panel and the stagnation point region in
the lower panel. Most of the $2-10 \kev$ X-ray emission outside
the X-ray minimum comes from the shocked secondary wind (e.g.,
Pittard \& Corcoran 2002). When the two stars approach periastron,
the ratio of the accretion radius of the secondary $R_{\rm acc2}$,
to the distance of the secondary from the stagnation point
$D_{g2}$, increases substantially, reaching a value of $\sim 0.1$,
40 days before periastron passage, and $\sim 0.4$ at periastron.
At the same time the free fall time of cold post-shock primary
wind mass elements becomes shorter, and it is no longer much
larger than the flow time out of the stagnation region. As pointed
out by Soker (2005b), the evolution of these two ratios strongly
suggests that the stagnation point region collapses onto the
secondary $\sim 0-40$~days before periastron passage, and the
secondary starts to accrete the primary wind. This process is
assumed to prevent the secondary from accelerating its wind, hence
shutting-down the main source of X-ray emission, causing the
 minimum.  The gradual collapse of the winds is associated
with a rise in density and column density, which is responsible
for the temporary hardening of the X-ray spectrum before minimum
(Corcoran 2005).

As distance increases after periastron passage the accretion rate
decreases, allowing the acceleration zone of the secondary wind
to build up again $\sim 60-70$ days after periastron passage.
As a result the wind reappears and so does the X-ray emission.
Right after the minimum, the binary separation is
still small and X-rays are emitted from deep in the primary wind,
again through a relatively high column density, which again
results in a (temporarily) hard spectrum.
The progressive hardening and softening of the X-ray spectrum just
outside the deep minimum is consistent with an absorption effect.
Later on, as the secondary emerges from
periastron, the radiation becomes softer returning to its normal
orbital spectrum.

{{{ At $\sim 30$~days into the  minimum the X-ray radiation
becomes harder for $\sim 25$ days. This is accompanied by an
increase in X-ray luminosity; this is observed in both the 1997.9
and the 2003.5 minima (Corcoran 2005).
}}} The small rise in X-ray luminosity $\sim 30$~days into the
minimum, {{{ might be explained with our model in the following,
somewhat speculative, way. }}} For a short time during the
accretion phase the specific angular momentum of the accreted mass
rises, such that the accretion onto the secondary is concentrated
in the equatorial plane, and part of the accreted mass is blown
along the polar directions. This moderate-velocity polar outflow
runs into the ambient primary wind and is subsequently shocked,
which results in weak extra X-ray emission during the  minimum.

Based on the results of Soker (2005b) and observations, we consider
the X-ray minimum to start $\sim 0-30$~days (phase $-0.015$ to
$0$) before periastron passage. In the present paper we take the
accretion to start 20 days before periastron (orbital-phase
$-0.01$), and to end 60 days after periastron, (orbital-phase
$0.03$). The X-ray minimum itself starts a little after accretion
starts, and lasts several days less than the assumed 80~day
accretion period. Any starting phase of the X-ray minimum in the
range $\sim -0.04-0$ is acceptable in our model.

\section{COLLIDING WINDS}
\label{sec:winds}

\subsection{Structure of the Flow}
\subsubsection{Binary orbit}

The binary system parameters are as in the first paper in this
series (Soker 2005b). The stellar masses are $M_1=120 M_\odot$,
$M_2=30 M_\odot$, the eccentricity is $e=0.9$, and orbital period
2024 days, hence the semi-major axis is $a=16.64 \AU$ and
periastron occurs at $r=1.66 \AU$. The mass loss rates are $\dot
M_1=3 \times 10^{-4} M_\odot \yr^{-1}$ and $\dot M_2 =10^{-5}
M_\odot \yr^{-1}$. {{{ Ishibashi (2001) proposed that $M_2
\simeq 40 M_\odot$. A more massive secondary implies even more
gravity by the secondary, favoring accretion near periastron
passage even further. According to Smith et al. (2003) the total
primary mass loss rate is higher than $3 \times 10^{-4} M_\odot
\yr^{-1}$, but with a higher mass loss rate in the polar
directions and lower mass loss rate toward the equatorial plane.
Since most of the X-ray emission comes from regions near the
equatorial plane, we take the mass loss rate as quoted above. }}}
The primary wind velocity profile is:
\begin{equation}
v_1=500 [1-(0.4 \AU/r_1)] \km \s^{-1} \label{eqv1}
\end{equation}
\noindent where $r_1$ is the distance from the center of the
primary.
The secondary wind speed is taken to be $v_2=3000 \km \s^{-1}$.
The orbital separation $r$, the relative orbital velocity of the
two stars $v_{\rm orb}$, and the angle $\theta$ of the position of
the secondary relative to the semi-major axis during periastron (see
Figure \ref{forbit1}) are plotted in the first row of Figure
\ref{forbit2}.

In the second row, the thick line represents the typical time
$\tau_{f2}$ for the shocked secondary wind to flow out of the
shocked region (winds interaction zone), while the thin line
depicts the radiative cooling time of the shocked secondary wind
$\tau_{\rm cool2}$.
In the third row of Figure \ref{forbit2}, the
distance of the stagnation point from the secondary, $D_{g2}$, and
the Bondi-Hoyle accretion radius of the primary wind by the
secondary star $R_{\rm acc2}$, are drawn. In the fourth row the
velocity of the primary wind relative to the stagnation point
$v_{\rm wind1}$ is depicted by the thick line, while the thin line
represents the ratio of $\tau_{f2}/\tau_{\rm cool2}$. For more
detail on these quantities see Soker (2005b).

\subsubsection{Colliding wind geometry}

The colliding wind region is schematically drawn in Figure \ref{fwinds}.
The winds from the two stars produce two respective shock waves.
The shocked gas flows away from the stagnation point along the
contact discontinuity$-$the surface where the two momentum fluxes
exactly balance each other and which
separates the two post-shock flows. The gas is heated by the shock
waves to temperatures of $\sim 10^{7}-10^{8} \K$ and generates X-ray
emission. The distances $D_1$ and $D_2$ from the binary components
to the stagnation point are calculated from the equation
\begin{equation}
\rho_1 v_{\rm wind1}^2=\rho_2 v_2^2 \label{r1}.
\end{equation}
Here $\rho_1$ and $\rho_2$ are the pre-shock densities of the two
winds, $v_{\rm wind1}$ is the pre-shock speed of the primary wind
relative to the stagnation point (assumed to move with the
secondary), and $v_2$ is the pre-shock speed of the secondary wind.
All quantities are calculated at the stagnation point. Using
equation (\ref{r1}) and the relation $D_1+D_2=r$ where $r$ is the
distance between the stars, the distance $D_1$ and $D_2$ can be
found. When the effect of the gravity of the companion is
included, the newly calculated distance of the stagnation point to
the secondary, $D_{g2}$, decreases slightly. At periastron $D_{g2}
\simeq 0.8D_2$.

The asymptotic half opening angle $\phi_a$ of the contact
discontinuity, which is defined in Figure \ref{fwinds}, is (e.g.,
Eichler \& Usov, 1993)
\begin{equation}
\phi_a \sim 2.1 \left(1-\frac{\beta^{4/5}}{4}\right)\beta^{2/3}
\simeq 1 \simeq 60^\circ, \label{theta}
\end{equation}
where
\begin{equation}
\beta \equiv \left( \frac{\dot M_2 v_2}{\dot M_1 v_1}
\right)^{1/2}. \label{beta}
\end{equation}
For the parameters used here $\beta \simeq 0.4$.

\subsection{X-ray Emission}

Usov (1992) gives an expression for the X-ray luminosity $L_x$
later used by Ishibashi et al. (1999). However, this expression
cannot be applicable here, because if we substitute the veocities
and mass loss rates typical of \etc\ at an orbital separation of
$r<4 \AU$ the total X-ray emission is more than the total kinetic
power of the secondary wind.
{{{ This is indeed a problem in the model of
Falceta-Goncalves et al.\ (2005). }}}
We therefore need to derive a
different expression for the X-ray luminosity. The expected
increase in the intrinsic luminosity with orbital separation is
$1/r$ (Usov 1992), much more than that observed.
The more moderate increase of $L_x$ with decreasing orbital separation was
attributed by Ishibashi et al.\ (1999) and Corcoran et al.\ (2001) to an
increase in absorption accompanying the increase in X-ray intrinsic luminosity
with decreasing radius. We will assume the same.

Usov (1992) uses a cooling function $\Lambda \propto T^{1/2}$ that
is appropriate for hot gas at $T \ga 2 \times 10^7 \K$. Since the
shocked primary wind of \etc\ is much cooler, we need to use a
more appropriate cooling function. We define $F_{AB}(T_s)$ to be
the fraction of the X-ray flux emitted by the gas shocked to the
temperature $T_s$ in the range $A-B$ (in keV), out of the total
flux emitted between 0.01--100~keV. The X-ray luminosity between
$2-10 \kev$ depends strongly on $T_s$. Using the APEC plasma
database (Smith et al. 2001),
we have calculated $F_{AB}(T_s)$ for a range of temperatures
corresponding to pre-shock wind velocities (relative to the shock)
of $500 \la v \la 2000 \km \s^{-1}$. We then fitted $F_{AB}(T_s)$
with a parametric form of the bremsstrahlung emissivity function
to obtain:
\begin{equation}
F_{210}=0.65 \exp \left(-2.325\times10^7/T_s\right)= 0.65 \exp
\left[ -6.6 \left(\frac{v}{500 \km \s^{-1}}\right)^{-2} \right].
\label{f210}
\end{equation}

\subsubsection{Primary wind}

Close to the X-ray minimum, which occurs near periastron passage,
the shocked primary wind is very dense and its cooling time is
much shorter than the flow time (e.g., Pittard \& Corcoran 2002;
Soker 2003). Therefore, we assume that any post-shock primary wind
material cools instantaneously by emitting all of its thermal
energy in the form of radiation. The emitted spectrum is taken to
be that of gas at the post-shock temperature at the stagnation
point. Since the shock front is oblique, the shock velocity and
the temperature away from the stagnation point, in reality, are
lower. Moreover, as the shocked primary wind loses energy, its
temperature decreases. Consequently, on average, the spectrum is
typical of temperatures lower than the immediate post-shock
temperature.
{{{{{ For temperatures typical of the shocked primary wind,
a small decrease in the emitting gas temperature substantially
reduces its contribution in the band above $2 \kev$. }}}}}
Therefore, it should be clear that our treatment overestimates the
contribution of the primary wind to the X-ray emission in the
$2-10 \kev$ band.

 From the shape of the shock wave for our relevant parameters (section
3.1) we estimate that the primary's wind segments with $\phi \la 40
^\circ$ pass through a strong enough shock and heat to high enough
temperature to contribute to the X-ray emission in the $2-10 \kev$ band
(see Fig. \ref{fwinds}).
This implies that the rate of mass entering the strong shock region is
$k_1 \dot M_1 = (1-\cos 40^\circ) \dot M_1/2=0.1 \dot M_1$.
The pre-shock velocity of this mass is about equal to the
relative velocity of the primary wind to the stagnation point,
$v_{\rm wind1}$, and it is plotted in the fourth row of Figure \ref{forbit2}.
The calculation of $v_{\rm wind1}$ includes the wind velocity
relative to the primary and the relative orbital velocity between the two stars.
The acceleration of the primary wind is included by taking a lower
value of $v_1$ close to the primary star, which increases with
distance from the primary up to a terminal velocity at large
distances of $v_1=500 \km \s^{-1}$ (see Soker 2005b for more details).

The relevant velocity for the primary wind at the stagnation point
is $v_{\rm wind1} \simeq 500 \km \s^{-1}$ (see fourth row of Fig.~
\ref{forbit2}). For a pre-shock velocity of $v_{\rm wind1}=400,
~500$ and $600 \km \s^{-1}$, corresponding to post-shock
temperatures of $T_{s1}=2.2 \times 10^6$, $3.5 \times 10^6 \K$,
and $5 \times 10^6 \K$, we find $F_{210}=  2.1 \times 10^{-4}$,
$1.1 \times 10^{-3}$, and $5.5 \times 10^{-3}$, respectively. In
other words, only a small fraction of the energy is emitted in the
X-ray band. The contribution of the primary wind is, therefore:
\begin{equation}
L_{x1} \simeq  2.4 \times 10^{33} \left( \frac {k_1}{0.1}\right)
\left(\frac {F_{210}}{0.001}\right) \left( \frac{\dot M_1}{3
\times 10^{-4}M_\odot \yr^{-1}}\right) \left( \frac{v_{\rm
wind1}}{500 \km \s^{-1}}\right)^2 \erg \s^{-1}. \label{lx1}
\end{equation}
The X-ray luminosity $L_{x1}$ as a function of orbital phase is
plotted as a thin solid line in the upper row of Fig.~\ref{flumx}.
The dotted line in the same figure shows the X-ray flux in the
entire $0.01-10 \kev$, namely taking ${F_{210}}=1$ in equation
(\ref{lx1}) (note the different scaling there).
Considering photoelectric absorption
(mostly at 2--3~keV), the observed luminosity reduces to $L_{x1}
\la 10^{33} \erg \s^{-1}$ during most of the orbit, with possible
intrinsic contribution of up to $L_{x1} \sim 10^{34} \erg \s^{-1}$
as the system approaches periastron where $v_{\rm wind1}$ is high
(Fig. \ref{forbit2}). On the other hand, close to periastron,
X-ray absorption is at its peak. The most important point,
however, is that the primary wind is {{{{{ very slow, therefore
the post-shock temperature is low and most of the X-ray emission
is soft with negligible contribution above $\sim 3 \kev$ (Pittard
\& Corcoran 2002). Furthermore, the X-ray emission below $3 \kev$
is more strongly absorbed, }}}}} and one must conclude that during most
of the orbit the major fraction of the observed $2-10 \kev$ X-ray
emission comes from the secondary wind.
{{{{{ (We emphasize again: }}}}} the
primary wind does emit in the $2-3 \kev$ band, as seen in the
upper row of Figure \ref{flumx}, however absorption in this band
is high and most of this flux never reaches the observer.
In addition, for reasons stated at the beginning of this subsection,
the treatment here overestimates the contribution of the shocked
primary wind to the $2-10 \kev$ band. )

\subsubsection{Secondary wind}
The secondary wind is much faster ($\sim$~3000~km~s$^{-1}$) than
the primary wind and the relevant temperature range is therefore
$\sim 0.5-1.3 \times 10^8 \K$ for which we find an average value
of $F_{210}=0.46$. At the high temperatures and low densities
typical of the secondary wind, and in contrast with the primary
wind, the radiative cooling time of the post-shock secondary wind
material $\tau_{\rm cool2}$, is much {\it longer} than the flow
time out of the wind-collision region.
This implies that the shocked secondary wind region is large, because the
shock front is at a large distance from the contact discontinuity, e.g.,
as Pittard et al.\ (2002) simulate for the massive binary system WR~147.
This is schematically drawn in Fig.~\ref{fwinds}.

For the purpose of studying the X-ray minimum, it is adequate to
take the contribution of the secondary wind to the X-ray
luminosity as follows. We assume that about half of the mass blown
by the secondary star is shocked in a shock front perpendicular to
the wind velocity. We then assume that the {{{{ X-ray
luminosity is determined by how much of the thermal energy of the
shocked gas is radiatively emitted as X-rays before the gas cools
adiabatically. The radiative emission lasts for time scales of the
order of the radiative cooling time $\tau_{\rm cool2}$, while
adiabatic cooling takes place on the flow time scale $\tau_{f2}$.
Therefore, }}}} the total X-ray energy emitted is a fraction
$k_2 \tau_{f2}/\tau_{\rm cool2}$ of the thermal energy of the
post-shock gas. Here $\tau_{f2} \equiv D_{g2}/v_2$ is the
characteristic flow time of the shocked wind out of the
interaction region, where $D_{g2}$ is the distance of the
stagnation point from the secondary. The value of
$\tau_{f2}/\tau_{\rm cool2}$ is plotted in the last row of
Fig.~\ref{forbit2}. The contribution of the shocked secondary wind
material to the X-ray luminosity in the energy range $A-B$ can
therefore be written as:

\begin{equation}
L_{x2} = \frac{1}{4} \dot M_2 v_2^2 F_{AB}(T_s) \frac
{\tau_{f2}}{\tau_{\rm cool2}}  k_2 . \label{lx2}
\end{equation}

\noindent Substituting typical values in equation~(\ref{lx2})
gives for the 2--10~keV range:

\begin{equation}
L_{x2} = 6.5 \times 10^{34} \left( \frac {F_{210}}{0.46}\right)
\left( \frac{\dot M_2}{10^{-5} M_\odot \yr^{-1}} \right)
\left( \frac{v_2}{3000 \km \s^{-1}}\right)^2          
\left( \frac {\tau_{f2}/\tau_{\rm cool2}}{0.01}\right) k_2 \erg
\s^{-1}. \label{lx22}
\end{equation}

As shown by Pittard \& Corcoran (2002), Kelvin-Helmholtz
instability modes develop on the contact discontinuity. This
instability indirectly enhances the X-ray emission by two effects:
($i$) the large "tongues" slow-down the velocity of the shocked
secondary wind, increasing the effective outflow time of shocked
secondary wind near the contact discontinuity; and ($ii$) the
Kelvin-Helmholtz instability mixes hot, shocked secondary-wind gas
with cool, dense shocked primary-wind gas. This mixing causes a
small fraction of the shocked secondary wind to cool to low
temperatures releasing most of its energy (and not just a fraction
of $\tau_{f2}/\tau_{\rm cool2}$). These effects, as well as our
ignorance of the exact values of the secondary wind speed and mass
loss rate, are incorporated in the parameter $k_2 > 1$ which we
fit to match the observed X-ray luminosity.
The luminosity $L_{x2}$ according to equation (\ref{lx22}) with $k_2=2$
is plotted as the thick line in the upper row of Figure \ref{flumx}
where it can be seen to dominate $L_{x1}$.

\subsection{X-ray Absorption}
If the X-rays from \etc\ are observed through the primary wind,
absorption is inevitable, but its magnitude depends on the wind
parameters, flow parameters, and system orientation. Firstly, the
soft ($2-5 \kev$) and hard ($7-10 \kev$) bands are generally
emitted from different regions. The hard X-rays come mainly from
regions closer to the stagnation point, where the shock wave is
strong and post-shock temperature is $\sim 10^8 \K$. This region
is marked by capital bold-face letters `X' in Figure
{\ref{fabsorb}}. Away from the stagnation point, the gas cools
adiabatically and the X-rays become softer. Thus, we expect the
column density toward the hard X-ray regions to be somewhat higher
than that toward the soft X-ray region. Although the soft X-ray
regions are less obscured, absorption of soft X-rays is high
enough to explain the X-ray hardening effect just prior and just
after the X-ray minimum. As our main goal is to explain the X-ray
light curve, including the minimum, in the present work we will not
model the $2-5 \kev$ and $7-10 \kev$ bands separately, but treat
the entire 2--10~keV band together.

Secondly, for some orientations and some of the time, the X-ray
emission could be observed through the tenuous secondary wind
instead of through the bulk of the dense primary wind, which would
imply a lower column density. These orientations are represented
in Fig.~\ref{fabsorb} by the narrow arrows. For observer A in
Fig.~\ref{fabsorb}, this occurs after the X-ray minimum, while for
observer B this occurs before the X-ray minimum. {{{ The winds
collision model implies significant absorption of the X-ray
emission during the X-ray maximum just before the  X-ray
minimum. This rules out the orientation of observer B in
Fig.~\ref{fabsorb}. (We note that Falceta-Goncalves et al.\ (2005)
and Abraham et al. (2005) argue to the contrary; however, they
require the line of sight to be through the orbital plane, which
is in contradiction with the structure of the Homunculus from
which the orbital plane is deduced to be tilted by $\sim 48
^\circ$ from an edge-on view (Smith 2002).) }}} At other phases,
the X-ray emission is partially absorbed by the much higher column
density of the primary wind (wide arrows in Figure \ref{fabsorb}).
To estimate the importance of the geometry, we assume that our
line of sight is at $\sim 45^\circ$ to the orbital plane (Smith
2002). Let $\phi_a$ be the asymptotic half opening angle of the
contact discontinuity (Fig.~\ref{fwinds}). At an angle of
$45^\circ$, the projection of the half opening angle on the
orbital plane, $\phi_{45}$, is given by $\tan \phi_{45}=(\tan^2
\phi_a-1)^{1/2}$. For $\phi_a=45^\circ$, $60^\circ$, and
$70^\circ$, we find $\phi_{45}=0$, $54.7 ^\circ$ and $68.7^\circ$,
respectively. Namely, if $\phi_a \le 45^\circ$ we will always
observe the X-ray emitting gas through the primary wind. For the
typical parameters used here $\phi_a \simeq 60^\circ $. Hence,
$\phi_{45} \simeq 55^\circ$. Therefore, during a substantial
fraction of the orbital orbit we might be observing the X-ray
emission through the fast secondary wind (i.e., low column
density). We do not elaborate on this effect in the present paper,
{{{ because there are too many unconstrained parameters to
consider: }}} (1) There is no consensus yet on the orbital
orientation (e.g., Ishibashi 2001; Smith et al.\ 2004). {{{ (2)
The non-spherical mass loss geometry from the primary, with denser
material in the polar direction (Smith et al. 2003). (3) The cone
depicted in Fig.~\ref{fabsorb} actually has a spiral structure due
to the orbital motion. Namely, the winds interaction region is
winding around the binary system as it flows outward, like the
dust in the interacting winds binary system WR~98a (Monnier et al.
1999). This includes the cool, dense post-shocked primary wind
material. Hence, some absorption is expected. Each one of these
three effects introduces at least one free parameter. At this
stage of the development of the model, there is no point in
exploring this immense parameter space. Instead, we notice that
most of the absorption occurs by the primary wind in regions close
to the stagnation point, and use this to build a simple geometry
which includes the main features of the primary wind near the
stagnation point. }}}
{{{{{ For comparison, we also consider the absorption for the
orbital orientation proposed by Smith et al. (2004). }}}}}

{{{ Following the discussion of the previous paragraph, }}}
in the present calculations we take, {{{{{ as our basic geometrical structure, }}}}}
the momentary column density through the primary wind to be that to the stagnation point,
along a line of sight perpendicular to the line connecting the two stars.
Taking $y$ to be the coordinate perpendicular to the line connecting the
two stars, and using ${\dot M_1}= n \mu m_H 4 \pi r_1^2v_1$, where
$r_1^2=D_1^2+y^2$, this hydrogen column number density can be expressed as:
\begin{equation}
N_{H1}=0.43 k_g \int_0^\infty \frac {\dot M_1}{4 \pi
(D_1^2+y^2)v_1 \mu m_H} dy,
\label{intN1}
\end{equation}
where $\mu m_H$ is the mean mass per particle in a fully ionized gas,
$n$ is the total number density, $43 \%$ of which are hydrogen nuclei,
and $k_g$ is a factor that depends on the geometry and orientation.
It is used here to compensate for the many unknown parameters.
Performing the integral under the assumption of constant $v_1$,
and substituting typical values near periastron gives
\begin{equation}
N_{H1}= 1.3 \times 10^{24} 
\frac{\dot M_1}{3 \times 10^{-4} M_\odot \yr^{-1}}
\left(\frac{v_1}{500 \km \s^{-1}} \right)^{-1} \left(\frac{D_1}{1
\AU} \right)^{-1}  k_g \cm^{-2}.
\label{N1}
\end{equation}
At apastron the binary separation is $31.6 \AU$, and the stagnation
point is a distance $D_1=r-D_{g2} \simeq 22 \AU$ from the primary,
which from Eq.~{\ref{N1}} yields a column density at apastron of
$N_{H1p} \simeq 6 \times 10^{22} \cm^{-2}$.
Observations indicate a column density of $N_{H} \simeq 3 \times
10^{22} \cm^{-2}$ during most of the orbit (Ishibashi et al.\
1999; Corcoran et al.\ 2001). We therefore take $k_g=0.5$ in the
present work, although this value can be larger when the system
emerges from the X-ray minimum.

In the second row of Fig.~\ref{flumx}, we plot the calculated
X-ray transmission factor using the column density given by
equation (\ref{N1}) with $k_g=0.5$. We plot the transmission
separately for the soft 2--5~keV, hard 7--10~keV, and the entire
2--10~keV bands. In the third row of Fig.~\ref{flumx}, we plot the
X-ray luminosity (first row), but now including absorption. It can
be seen that after absorption $L_{x2}$ is even more dominant over
$L_{x1}$.

{{{{{  The geometry used to calculate the column density in
equations (\ref{intN1}, \ref{N1}) was the simplest one, and we chose it
because of the many uncertainties. We could use another geometry,
such as that suggested by Smith et al. (2004): The inclination angle
$i$ of our line of sight to a norm to the orbital plane
is $42^\circ$, and the secondary is on the far side of
the primary before reaching periastron (the semi-major axis
is perpendicular to our line of sight).
In this geometry, the line of sight to the stagnation point practically
always goes through the primary wind, unlike some orientations shown
in Fig.~\ref{fabsorb}, which was plotted for the $i= 90^\circ$ case.
The reason is that after periastron passage the bow shock of the secondary
wind is fully recovered only at phase $\sim 0.04$ (see fig. \ref{flm}), when the
orbital angle is $\theta \sim 140^\circ$ (upper row of Figure \ref{forbit2}).
It can be shown that when the semi-major axis is perpendicular to
the line of sight then the column density of the primary wind as a function
of the binary azimuthal angle $\theta$ and inclination angle $i$ is:
\begin{equation}
N_{H1i}(i)=\frac{0.43 k_g \dot M_1}{4 \pi \mu m_H v_1 D_1}
\frac{(1+ \tan^2 i)^{1/2}}{(1+\cos^2 \theta \tan^2 i)^{1/2}}
\left[  1-\frac{2}{\pi} \tan^{-1}
\frac{\sin \theta \tan i}{(1+\cos^2 \theta \tan^2 i)^{1/2}} \right]
\label{intNi}
\end{equation}
where $\theta$ here is negative before periastron, zero at periastron,
and positive after periastron (see Fig.~1).
The ratio $N_{H1i}(i=42^\circ)/N_{H1}$ as a function of the orbital
phase is plotted in the upper row of Fig.~\ref{fincline} for a fixed
value of $k_g$.
In the second row of Fig.~\ref{fincline}, the absorbed X-ray luminosity is plotted
with $k_g=0.5$,
but with the column density given by equation (\ref{intNi}).
The last row of Figure \ref{fincline} shows the ratio of this X-ray luminosity
to that shown in Fig. \ref{flumx} for the $i= 90^\circ$ case.
It can be seen that the only significant difference arises
near periastron, where according to the proposed model, the secondary wind does
not exist. We conclude that the difference between the two geometries is
practically small and that the X-ray luminosity is not very
sensitive to the inclination angle $i$ (as long as $i$ is not too small).
The emission of the shocked primary wind is reduced substantially; but
in our model most X-ray emission in the $2-10 \kev$ bend is attributed to the
shocked secondary wind anyway. }}}}}
{{{{{{ To summarize, although the binary orientation assumed in calculating the
absorbing column density used in equation (\ref{intN1}) (or \ref{N1}) is not the
exact orientation of the binary system in $\eta$~Car, it is adequate for the main goal
of the present paper as it emphasizes the weak dependence of the results on the exact
geometry.
The orientation used in deriving equation (\ref{intN1}) has the advantage that
it does not depend on the orientation of the periastron ($\omega$) with respect
to our line of sight.
The small difference between the results obtained for the two geometries used here
demonstrates that our model is neither sensitive to the orientation of the periastron
nor to the inclination of the orbital plane (as long as $i$ is not too small). }}}}}}

The X-ray minimum lasts $\sim 70$ days, the time from
the beginning of the decline to the end of the rise from the
minimum is $\sim 120$~days. According to our proposed model, the
decline starts $\sim 40$~day before periastron passage ($r \simeq
5.3 \AU$), when the absorbing column density increases and blobs
from the post-shock primary wind start to be accreted. Bondi-Hoyle
type accretion starts $\sim 20$~days before periastron ($r \simeq
3.3 \AU$). The  minimum starts $\sim 10$~days before
periastron, and ends $ \sim 60$ days after periastron ($r \simeq
7.2 \AU$), taking another $\sim 20$~days to fully rise back
($r\simeq 8.7 \AU$, and $D_1\simeq 6 \AU$). At the beginning of
the decline (40 days before periastron), where the absorption
effect is most significant, the column density is $N_{H1} \simeq
1.5 \times 10^{23} \cm^{-2}$ (with $k_g=0.5$). At this column
density, 29\%, 83\%, and 71\% of the flux is transmitted
through the primary wind in the $2-5 \kev$, $7-10 \kev$, and
2--10~keV bands, respectively. The rest is absorbed by the primary
wind.

\begin{table}

Table 1: Glossary

\bigskip
\begin{tabular}{|l|l|c|}
\hline
Symbol  & Meaning   & Typical value \\
\hline \hline
$r$ & Binary separation& $1.66-31.6 \AU$  \\
\hline
$\theta$ & Binary azimuthal angle (Fig.~1)& $0 - 2\pi$  \\
\hline
$D_1$& Distance of stagnation point from primary
$D_1=r-D_{g2}$ & $\sim 0.75~ r$
\\
\hline $D_2$& Distance of stagnation point from secondary; gravity
neglected & $\sim
0.3~ r$ \\
\hline $D_{g2}$ & Distance of stagnation point from secondary;
$M_2$ gravity included &
$\sim 0.25~ r$ \\
\hline
$M_1$& Mass of primary   & $120 M_\odot$ \\
\hline
$M_2$& Mass of secondary   & $30 M_\odot$ \\
\hline
$\dot M_1$& Mass loss rate by primary   & $3 \times 10^{-4} M_\odot \yr^{-1}$ \\
\hline
$\dot M_2$& Mass loss rate by secondary & $10^{-5} M_\odot \yr^{-1}$ \\
\hline
$\dot M_{\rm acc2}$ & Mass accretion rate of primary wind by secondary & [eq. {\ref{macc2}}]$_a$ \\
\hline
$v_1$ & Velocity of primary wind (relative to primary)  &  $500 \km \s^{-1}$ \\
\hline
$v_2$ & Velocity of secondary wind (relative to secondary)
& $3000 \km \s^{-1}$
\\
\hline ${v_{\rm wind1}}$ & Velocity of primary wind relative to
secondary  &  $\sim
500 \km \s^{-1}$ \\
\hline
$R_2$ & Stellar radius of secondary& $\sim 0.1 \AU$  \\
\hline
$R_{\rm acc2}$  & Accretion radius of secondary & $\sim 0.2-0.4 \AU$ \\
\hline
$\tau_{\rm cool2}$ & Radiative cooling time of post shock secondary wind & $[$5 days$]_p$\\
\hline
$\tau_{f2}$ & Flow time of post shock secondary wind & $[$0.2 days$]_p$ \\
\hline
$F_{210}$ & Fraction of the X-ray energy emitted in the $2-10 \kev$ band & 0-1\\
\hline
$L_{x1}$& Unabsorbed X-ray luminosity of the shocked primary wind& eq. (\ref{lx1})  \\
\hline
$L_{x2}$& Unabsorbed X-ray luminosity of the shocked secondary wind   & eq. (\ref{lx2})  \\
\hline
$L_{xa1}$& Unabsorbed X-ray luminosity of accreted primary wind  & [eq. (\ref{lxa1})]$_a$  \\
\hline
$L_{xa2}$& Unabsorbed X-ray luminosity of primary wind hitting the secondary  & [eq. (\ref{lxa2})]$_a$  \\
\hline
$N_{H1}$ & Hydrogen column density of obscuring primary wind& $\sim 10^{23} \cm^{-2}$  \\
\hline
$k_1$& Fraction of primary wind that is strongly shocked & $\sim 0.1$ \\
\hline
$k_2$& Parameter adjusting the X-ray luminosity of secondary wind & $\sim 2$ \\
\hline
$k_g$& Parameter adjusting column density of obscuring primary wind & $\sim 0.5-1$ \\
\hline
$i$ & Inclination angle (of line of sight to the norm of the orbital plane) & $0 - \pi / 2$  \\
\hline
$v_{c1}$&Pre-shock velocity of the primary wind on the accretion line & [eq. {\ref{vc1}}]$_a$ \\
\hline
$v_{ct1}$&Tangential component of $v_{c1}$ & [eq. {\ref{vct1}}]$_a$ \\
\hline
$d \dot m/dz$ & Rate of mass hitting the accretion column
per unit length & [eq. {\ref{dmdz}}]$_a$ \\
\hline
z & Distance from secondary on the accretion line (Fig. \ref{facc}) &\\
\hline
$z_{\rm min}$ & Minimum distance of accretion shock from secondary & [$0.1 \AU$]$_a$\\
\hline
\end{tabular}

\footnotesize
\bigskip

Notes: $[\ ]_p$ indicates values given at
periastron.\\
$[\ ]_a$ indicates values defined only for the $\sim 80$~days
accretion period.

\normalsize
\end{table}
\section{ACCRETION PHASE}
\label{sec:time}
\subsection{Flow Structure}

In the presently proposed model, the X-ray minimum is assumed to
occur as the secondary accretes mass from the primary wind (Soker 2005b).
The accreted mass is assumed to prevent the secondary from blowing its
fast wind, thus the dominant X-ray source is turned off. The
accretion flow is of the Bondi-Hoyle type (Figure \ref{facc}). The
very high Mach number creates an accretion shock. Because the
cooling time of the post-shocked primary wind material is very
short near periastron, about one percent of the flow time, (e.g.,
Pittard \& Corcoran 2002; Soker 2003), the accretion flow can be
treated as isothermal. Such a flow was simulated, e.g., by Ruffert
(1996). In the following, we distinguish between segments of the
primary wind impacting the secondary directly and segments that go
around the secondary toward the accretion column (behind the
secondary) along a curved trajectory. The basic flow structure is
as follows.

The accretion column$-$the elongated region behind the secondary
enclosed by the shocked accreted matter$-$is narrow (see Fig. \ref{facc})
and the Mach number of the pre-shock primary wind is high.
As an approximation, we take the velocity of the primary
wind just before it hits the shock wave as the velocity it would
have reached on the accretion line$-$the symmetry line behind the
accreting body. Using the definition of $R_{\rm acc2}$
\begin{equation}
R_{\rm acc2}= \frac {2 G M_2}{v_{\rm wind1}^{2}}  = 0.2   
\left( \frac{M_2}{ 30 M_\odot}\right) \left( \frac {v_{\rm
wind1}}{500 \km \s^{-1}} \right)^{-2} {\rm AU}, \label{accrad}
\end{equation}
this velocity can be found from energy conservation to be:
\begin{equation}
v_{c1}=v_{\rm wind1} \left(1+\frac{R_{\rm acc2}}{z} \right)^{1/2}
\label{vc1}
\end{equation}
where $z$ is the distance from the secondary along the accretion line.
The component in the perpendicular direction with respect to the
accretion line of the primary-wind velocity before hitting the accretion-column
shock $v_{ct1}(z)$, is found from angular momentum conservation.
Assuming the stream lines of the primary wind to be parallel at large distances
from the secondary, we can write
\begin{equation}
v_{ct1}=v_{\rm wind1} \left(\frac{R_{\rm acc2}}{z} \right)^{1/2}.
\label{vct1}
\end{equation}
Simulations of high Mach number, nearly-isothermal flows (e.g.,
Ruffert 1996) show that the flow structure is very similar to that
of the classical Bondi \& Hoyle (1944) accretion flow. The total
accreted mass can be approximated for our purposes as follows:
\begin{equation}
\dot M_{\rm acc2} \simeq  \pi \rho_0 R_{\rm acc2}^2 v_0 = 0.01
\dot M_1 \left(\frac {r}{1 \AU} \right)^{-2}
  \left(\frac {R_{\rm acc2}}{0.2 \AU} \right)^{2}
  \left(\frac {v_1}{500 \km \s^{-1}} \right)^{-1}
  \left(\frac {v_{\rm wind1}}{500 \km \s^{-1}} \right),
\label{macc2}
\end{equation}
where we take the density far upstream to be equal to the density
at the location of the secondary $\rho_0=\rho_1(r)=\dot M_1/(4 \pi
r^2 v_1)$, and the upstream speed is $v_0=v_{\rm wind1}$.
The mass accretion rate as function of the orbital phase is plotted in the
upper panel of Figure \ref{flacc}, only during the orbital phases
where according to our model accretion occurs. Part of the
accreted mass hits the accretion line in the range $z_{\rm min}
\la z \la R_{\rm acc2}$, while another fraction of the incoming
mass, $\sim (z_{\rm min}/R_{\rm acc2}) \dot M_{\rm acc2}$, hits
the accretion shock close to the accreting body. The rate of mass
hitting the accretion column per unit length is (Bondi \& Hoyle 1944)
\begin{equation}
\frac {d \dot m_z}{dz} =\pi \rho_1(r) v_{\rm wind1} R_{\rm acc2}.
\label{dmdz}
\end{equation}

\subsection{X-ray Emission due to Accretion}
The material with impact parameter $b \la R_{\rm acc2}$ will be attracted by
the secondary's gravity and flow towards the accretion line,
passing through the shock almost perpendicular to the shock front.
This gas is subsequently heated to a post-shock temperature appropriate
for its incoming velocity, which is given by equation (\ref{vc1}).
We assume this gas to be completely thermalized.
The thermal energy gained per unit length is:
\begin{equation}
\frac {d \dot E_{a1}}{dz} = \frac{1}{2} \frac {d \dot m_z}{dz}
v_{c1}^2. \label{ea1}
\end{equation}
We further assume that the cooling time is very short and that all
this energy is radiated away giving rise to an X-ray luminosity in
the $2-10 \kev$ range of:
\begin{equation}
L_{xa1} = \frac{1}{2} \int_{z_{\rm min}}^{R_{\rm acc2}} F_{210}(z)
\frac {d \dot m_x}{dz} v_{c1}^2 dz, \label{lxa1}
\end{equation}
where $F_{210}$ depends on $z$ through the dependence of the
post-shock temperature $T_s$ on $v_{c1}$.

The material impacting the secondary directly is also assumed to pass
through the accretion shock perpendicular to the shock front at a
distance $z_{\rm min}$ from the secondary, with a speed given by
equation (\ref{vc1}) with $z=z_{\rm min}$.
We take $z_{\rm min}$ to be approximately the
radius of the secondary star $R_2 \simeq 22 R_\odot \simeq 0.1 \AU
$ (Verner et al. 2005). This contribution to the X-ray luminosity
is therefore
\begin{equation}
L_{xa2} = \frac{1}{2} \frac{z_{\rm min}}{R_{\rm acc2}} \dot M_{\rm
acc2} \left( F_{210} v_{c1}^2 \right)_{z=z_{\rm min}}.
\label{lxa2}
\end{equation}
For $z_{\rm min}=0.1 \AU$ and near periastron we find for the
pre-shock speed $v_{c1} \simeq 900 \km \s^{-1}$, and for the
post-shock temperature $T_s \simeq 10^7 \K$ and $F_{210} \simeq
0.08$. We scale equation (\ref{lxa2}) to find
\begin{equation}
L_{xa2} \simeq 4 \times 10^{34} \left( \frac {z_{\rm min}/R_{\rm
acc2}}{0.5} \right) \left( \frac{\dot M_{\rm acc2}}{3 \times
10^{-6} M_\odot \yr^{-1}} \right) \left( \frac{F_{210}}{0.08}
\right) \left( \frac{v_{c1}}{1000 \km  \s^{-1}} \right)^2 \erg
\s^{-1}. \label{lxa2s}
\end{equation}
{{{ The two contributions to the intrinsic luminosity in the
$2-10 \kev$ band are plotted in the second panel of Figure
\ref{flacc}. The third panel shows the total luminosity of the two
contributions, namely, in the $0-\infty \kev$ band. }}}
This intrinsic luminosity is
much larger than the luminosity observed during the  minimum.
However, during the minimum, the column density through the
primary wind is considerable ($N_{H1} > 10^{23} \cm^{-2}$) and it
totally absorbs the X-ray accretion source. We find that accretion
cannot even account for the weak X-ray flux during the
minimum as demonstrated below.

The mass hitting the accretion column at $z \ga {R_{\rm acc2}}$,
is not accreted by the secondary. It passes
through an oblique shock at an angle of $\theta$ between the
inflow direction and the shock front with $\tan \theta \simeq
v_{ct1}/v_{\rm wind1}$ (or $\sin \theta = v_{ct1}/v_{c1}$).
This post-shock temperature is less than the post-shock
temperature of the primary wind when accretion does not occur.
Hence this gas emits soft X-rays, which are totally absorbed by the
primary wind. We do not consider its contribution in this paper.

As the primary wind flows toward the accreting body its density
increases. Therefore, the column density to the X-ray emitting
regions here is somewhat larger than that in equation (\ref{N1}).
In addition, the density in the accretion column is very large
making it completely opaque. Therefore, only $\sim$~half of the
intrinsic luminosity can be observed. For the emerging radiation,
we take the column density of equation (\ref{N1}), but with $r
\simeq 2 \AU$ replacing $D_1$, and a lower value for $v_1$
\begin{equation}
N_{H1}= 10^{24} 
\frac{\dot M_1}{3 \times 10^{-4} M_\odot \yr^{-1}}
\left(\frac{v_1}{400 \km \s^{-1}} \right)^{-1} \left(\frac{r}{2
\AU} \right)^{-1} \cm^{-1}. \label{Na1}
\end{equation}

The expected X-ray emission given by equations ({\ref{lxa1}}) and
({\ref{lxa2s}}), including absorption (eq.~{\ref{Na1}}) is plotted
in the lower panel of Figure \ref{flacc}. This radiation is soft,
with most of the contribution to the $2-10 \kev$ band coming from
radiation in the $2-3.5 \kev$ energy range.
These simplified estimates clearly show that the
soft emission due to accretion together with the high column
density it needs to traverse rules out this component as a
candidate for the weak X-ray emission during the  minimum. A
different alternative is pursued in the next subsection.

\subsection{X-ray Emission during the  Minimum}

{{{{ The main goal of the present paper is to show that the
X-ray behavior up to the minimum can be explained by the assumed
orbital parameters and by the variation of the radiative cooling
timescale of the secondary's wind relative to its flow timescale.
}}}} The present subsection, on the other hand, is more
speculative. We put forward a suggestion for the nature of the
very weak X-ray emission during the  minimum. We already showed
that X-ray emission due to accretion cannot explain even the weak
X-rays observed during the minimum. Because of the speculative
nature of this subsection, the treatment is more qualitative. We
suggest that the weak X-ray emission during the minimum, with its
variation in hardness and intensity (see property 3 in section
2.1), could come from two different components: ($i$) an almost
constant weak contribution from old shocked secondary wind, termed
here the residual component; and ($ii$) a possible temporary
collimated outflow resulting from accretion of mass with
relatively high specific angular momentum. These two components
are treated in the next two subsections.

{{{ We do not attempt to refute other contributions, e.g.,
other stars in the field, and more important the possibility that
most of the X-ray emission during the X-ray minimum comes from
reflected light by the nebula (Hamaguchi et al. 2005); time delay
ensures that the X-ray emission is observed during the minimum (Corcoran et
al. 2004a). We only point out that other processes should be
considered as well, before further observations and calculations
will reveal the most significant process contributing to the X-ray
emission and its variability during the  minimum. }}}

\subsubsection{Residual emission}
\label{sec:res}
We propose that {{{ some fraction of }}}
the soft and weak X-ray emission observed even
during the  minimum results from previously shocked secondary
wind segments. This residual emission exists basically along the
entire orbit, but it can be noticed only after the main X-ray
source has been shut down.

During the winds collision phase, the shocked secondary wind forms
a hot bubble flowing away from the secondary in what we term the
secondary hot-tail, winding around the binary system as it flows
outward, like the dust in the interacting winds binary system
WR~98a (Monnier et al. 1999). The shocked secondary wind is
engulfed eventually by the slower primary wind, and expands with
it to large distances at a speed of $v_1$. We wish to crudely
estimate the contribution of such a bubble to the X-ray emission.
This crude estimate is by no means a replacement for a full
3-dimensional numerical simulation of this flow, but it does give
a rough idea of its expected X-ray luminosity.

Let the main contribution to the residual emission come from
gas shocked during an average time $t_r$ before present, where
$t_r$ is a substantial fraction of the orbital period. Let a
fraction $q_m$ of the mass blown by the secondary during this
time, $q_m \dot M_2 t_r$, be enclosed in a fraction $q_V$ of the
volume $4 \pi (v_1 t_r)^3/3$. Because of the expansion, this gas
is much cooler than the post strong-shock temperature of $\sim
10^8 \K$.
Say $T(t_r) \sim 10^7$~K, for which the
contribution of the cooling gas to the emission in the $2-10 \kev$
band is only $F_{210} \sim 0.06$ (eq. \ref{f210}).
Because of the expansion to large distances, this emission is hardly absorbed
in comparison to the case for the accretion X-rays, although some absorption still
exists. Taking gas at $\sim 10^7$~K, we find the X-ray emission to be:
\begin{equation}
L_{x {\rm {-res}}} \sim 10^{34} 
\frac{q_m^2}{q_V} \left( \frac {F_{210}}{0.05}\right)
\left(\frac{t_r}{1 \yr} \right)^{-1} \left( \frac{\dot
M_2}{10^{-5} M_\odot \yr^{-1}} \right)^2 \left(\frac{v_1}{500 \km
\s^{-1}} \right)^{-3} \erg \s^{-1}.
\label{lres}
\end{equation}
This suggests that old shocked secondary wind might in principle
account for {{{ a significant fraction of }}}
the soft $\sim 3 \times 10^{33} \erg \s^{-1}$ emission during
the  X-ray minimum. In a future paper the
spectrum of this radiation will be compared with observation.

\subsubsection{Collimated outflow during the accretion phase}
\label{sec:cfw}
We consider the specific angular momentum of the accreted matter
$j_a$, and compare it to $j_2=(G M_2 R_2)^{1/2}$, the specific
angular momentum of a particle in a Keplerian orbit at the equator
of the accreting star of radius $R_2$. When a compact secondary
star moves in a circular orbit and accretes from the wind of a
mass losing star, such that the accretion flow reaches a steady
state, the ratio of the specific angular momenta is (Soker 2001)
\begin{equation}
1< \frac {j_a}{j_2} \simeq 0.1 ~ \eta \left( \frac {M_1+M_2}{150
M_\odot} \right)^{1/2} \left( \frac {M_2}{30 M_\odot}
\right)^{3/2} \left( \frac {R_2}{20 R_\odot} \right)^{-1/2} \left(
\frac {r}{2 \AU} \right)^{-3/2} \left( \frac{v_{\rm wind1}}{400
\km \s^{-1}} \right)^{-4},
\label{jacc}
\end{equation}
where $\eta$ is the ratio of the accreted angular momentum to that
entering the Bondi Hoyle accretion cylinder.
There is a net accreted angular momentum because more mass
is accreted by the secondary from the denser region facing the
primary than from the other side, behind the secondary.
When a steady state accretion flow is reached, the accretion column
(and the accretion line) bends toward the lower density region,
partially compensating for its lower density, and reducing the net
accreted angular momentum.
For this case numerical simulations show that $\eta \sim 0.2$
(e.g., Ruffert 1999).
However, here no steady state is
reached around phase $\sim 0.004$, when the accretion radius increases
by a large factor in a short time (Figure \ref{forbit2}). This
increases the mass accretion rate in a short time (Figure
\ref{flacc}) and therefore the accretion column has no time to
bend and reduce the specific angular momentum of the accreted gas.
This is the reason for taking $\eta=1$ here.

As seen, at phase $\sim 0.004$, which occurs $\sim 8~$days after
periastron, the specific angular momentum is quite large, although
still less than that required to form an accretion disk
($j_a/j_2>1$). This implies that the accreted mass will be
concentrated near the equatorial plane, with lower density regions
along the polar directions.
Consequently, the strong wind blown by the secondary
(when undisturbed) is not efficiently suppressed along the polar
directions, and together with a disk wind from the accreted matter
lead to the formation of jets or a biconical collimated wind.
This wind is then shocked by the impact of the ambient
primary wind, and emits the low-level X-rays observed in the
middle of the  minimum. This emission is highly obscured by
the high column density of the primary wind, hence it contributes
mostly in the hard band. We show that the X-ray emission during
minimum can be fitted by reasonable parameters. Let us take the
polar outflow to be at a speed of $v_p=2000 \km \s^{-1}$, with the
mass outflow rate being a fraction of $0.05$ of the accreted mass.
Taking the accretion rate at phase $0.05$ (10 days after periastron passage)
from our calculations (upper panel of Figure \ref{flacc}) to be
$\sim 10^{-6} M_\odot \yr^{-1}$, the total kinetic energy in this
polar outflow is $L_{\rm p-kin} \sim 6 \times 10^{34} \erg s^{-1}$.
For a column density of $10^{24} \cm^{-2}$, all the soft
X-ray is absorbed while the hard X-rays, comprising $\sim 0.4$ of
the flux, is reduced by a factor of $\sim 5$, giving a luminosity
of $L_{p-x}\simeq 5 \times 10^{33} \erg s^{-1}$. This hints that a
short-duration collimated outflow might account for the increase
in the middle of the  minimum.
{{{ On even shorter time scales and smaller amplitudes, stochastic variations
in the accretion process$-$both mass accretion rate and angular momentum accretion
rate$-$which are well documented in numerical simulations (Ruffert 1996),
might lead to the small spikes observed during the  minimum. }}}

\section{DISCUSSION AND SUMMARY}
\label{sec:summary}

As was shown by Soker (2005b), the effect of the gravity of the
secondary star on the primary wind becomes significant as the
system approaches periastron.  Cold and dense blobs are likely to
form in the post-shock primary-wind region, which becomes
unstable.  It seems plausible that close to periastron passage,
when the secondary's gravity becomes significant, these dense
blobs will be accreted by the secondary. Adopting the hypothesis
of Soker (2005b), we have {\it assumed} that these segments of the
primary wind, which are accreted, vigorously disrupt the
acceleration zone of the secondary wind, so that the secondary
wind ceases to exist.
The formation of cold blobs will be verified in a future paper
via 3D-gasdynamical numerical simulations.
The effect of binary accretion on the launch of stellar winds will have
to be studied theoretically in the future by examining the sensitivity
of the wind acceleration zone in O stars to accreted cold gas.

In the present paper, the accretion of dense blobs is assumed to
start $\sim 40-50$~days before periastron passage (orbital phase
$(\sim -0.025)- \sim (-0.02)$), and the Bondi-Hoyle type accretion
flow (Fig. \ref{facc}) is assumed to start $\sim 20$~days before
periastron passage (orbital phase $\sim -0.01$); several days
later, the (almost flat) X-ray minimum starts. As can be seen from the
upper panel of Figure \ref{flacc}, $\sim 60$ days after periastron
passage (orbital phase $\sim +0.03$) the accretion rate diminishes
and it is {\it assumed} that the secondary star builds back its
acceleration zone and its wind reappears. This is when the system
starts to get out of the X-ray minimum, a process lasting $\sim
20$~days.
%

An interesting feature of the proposed model is the asymmetry
around periastron in the relevant properties. The asymmetry
results from the asymmetry of the relative velocity of the primary
and secondary winds at the stagnation point. As the two stars
approach each other, the relative velocity is higher than when
they recede. This can be seen in the lower row of Figure
\ref{forbit2}. This effect causes the stagnation distance
$D_{g2}$, the accretion radius $R_{\rm acc2}$, the post-shock
primary wind temperature, and other quantities to acquire
asymmetric values about periastron passage.

{{{ In the proposed model the steep decline in the X-ray emission
is due to the collapse of the stagnation-point region onto the
secondary, very close to periastron passage (Soker 2005b). The
collapse starts a few weeks before periastron passage, {{{{
when the secondary's gravity at the stagnation point becomes
significant. This occurs when the ratio of the accretion radius to
the stagnation distance $R_{\rm acc2}/D_{g2}$ increases to
$\sim$~few~$\times 0.1$. For the assumed wind parameters in the
$\eta$ Car system, when this occurs, the ratio of the outflow time
of the shocked primary wind, $\tau_{f1}$, to the free fall time
$\tau_{ff2}$, becomes $\sim$~few~$\times 0.1$ as well, further
supporting the importance of accretion. At present, we can not say
what exact value of these ratios is needed for accretion to start.
This would require 3D hydrodynamical numerical simulations.
Furthermore, there are large uncertainties in the binary and wind
parameters of $\eta$ Car. Phenomenologically, what we can say is
that the accretion model can account for the behavior around the
X-ray minimum of $\eta$ Car if accretion starts when
$R_{\rm acc2}/D_{g2} \simeq 0.2$  (Soker 2005b). This occurs when the
stagnation point distance from the secondary is
$D_{g2} \simeq 0.6 \AU$.
We therefore assume that accretion starts at that phase.
}}}} The collapse occurs over a time period of a few times the
free-fall time from the stagnation point to the secondary
\begin{equation}
\tau_{\rm ff2}= 5.5  
\left( \frac{M_2}{30 M_\odot}\right)^{-1/2}
\left( \frac {D_{g2}}{0.6 \AU} \right)^{3/2} ~{\rm day}.
\label{freef}
\end{equation}
The details of this process requires 3D numerical simulations, which are beyond
the scope of the present paper. We therefore make do with a simple
phenomenological approach.
{{{{{ The start of the minimum is assumed to be the time when $\tau_{f1}/\tau_{ff2}$
and $R_{\rm acc2}/D_{g2}$ reach some critical value. From the X-ray
lightcurve we assume the minimum starts when
$\tau_{f1}/\tau_{ff2}$ and $R_{\rm acc2}/D_{g2}$ reach $\sim 0.25$
and $\sim 0.2$, respectively.
We assume that the minimum ends when these ratios fall back
to these same values.
Values of $\tau_{f1}/\tau_{ff2} \simeq 0.25$ and $R_{\rm acc2}/D_{g2}\sim 0.2$
yields a duration of $\sim 70$ days for the minimum
(Soker 2005b), in good agreement with observations.
We assume a simple linear approximation for the decline
to minimum and subsequent recovery.  }}}}}
We take the significant contributions to the total X-ray
luminosity $L_{210}$ in our simple model, to be
\begin{eqnarray}
L_{210}(p) =  L_{x {\rm {-res}}} + \left\{
\begin{array}{cl}
L_{x1}+L_{x2}  & \qquad p<-0.015 \quad {\rm or } \quad p>0.04 \\
(-0.005-p)(L_{x1}+L_{x2})/0.01  & \qquad -0.015 \le p < -0.005  \\
0  & \qquad -0.005 \le p < 0.03  \\
(p-0.03)(L_{x1}+L_{x2})/0.01  & \qquad 0.03 \le p \le  0.04,
\end{array}
 \right .
\label{lm}
\end{eqnarray}
where the phase $p$ is defined in the range $-0.5 \le p < 0.5$.
We have included $L_{x1}$ and $L_{x2}$ from
equations (\ref{lx1}) and (\ref{lx22}), but we ignore, for the
present purpose, the small contributions of accretion and the
residual emission mechanisms during the minimum (sections
\ref{sec:res} and \ref{sec:cfw}). In order to compare with
observations, we need to consider absorption. The absorbed X-ray
luminosity, namely, $L_{210}$ but with the absorption taken into
account is given in Figure \ref{flumx}. The final X-ray luminosity
predicted by our model is drawn in Figure \ref{flm} together with
the data from Corcoran (2005).

{{{{{
The binary system's parameters were chosen here from the literature on
$\eta$ Car, and following Soker (2005b).
We note that for different parameters, the proposed accretion scenario
might even be more plausible.
Changing the secondary mass from $M_2=30 M_\odot$ to $M_2=40 M_\odot$
(Ishibashi 2001), and keeping the eccentricity and orbital
period unchanged, will not affect much the periastron distance.
However, the accretion radius will increase by a factor of $\sim 1.3$,
making accretion more likely.
However, to make the ratio $R_{\rm acc2}/D_{g2} >0.5$ at phase $\sim -0.01$
(when accretion starts) will require the secondary mass to be $M_2>60 M_\odot$,
which seems unlikely.

More likely to influence this ratio is a lower primary wind speed near
periastron.
If the wind speed is low, say 30\% of the 500~km~s$^{-1}$ assumed here near periastron,
then the relative velocity of the secondary to the primary wind will
basically be the orbital speed, $\sim 250 \km \s^{-1}$ at phase $-0.01$,
instead of $\sim 500 \km \s^{-1}$.
The distance to the stagnation point $D_{g2}$ will not change much, because the
lower speed causes an increase in the primary wind density.
The accretion radius will be $\sim 4$ times its
value in the present case, bringing the ratio $R_{\rm acc2}/D_{g2}$ to about unity.
Moreover, the slower primary wind will increase the outflow time
of the postshock gas $\tau_{f1}$, by that increasing the ratio $\tau_{f1}/\tau_{ff2}$.
The slower primary wind speed may result from enhanced mass loss rate as a result
of the influence of the companion, or from extended wind acceleration zone.
The point is that parameters might be more favorable for the accretion scenario
than those used by Soker (2005b), but not by much. }}}}}

We do not try to match the observation point by point, but rather
to catch the general behavior with emphasize on the X-ray
minimum. We use $k_2=2$ to match the observation near apastron,
when attenuation is weak. We then find that the luminosity before
and after minimum can be matched quite well with column density as
given by equation (\ref{N1}) with $k_g=0.5-1$. In the upper row on
Figure  \ref{flm} we draw the expected X-ray luminosity in the
$2-10 \kev$ band according to our model for $k_g=0.5$ (thick line)
and $k_g=1$ (thin line). In the lower row the $k_g=0.5$ case is compared
with the X-ray flux covering two spectroscopic events (two thin lines).
By playing more with the several parameters, like the winds properties
and eccentricity, an even better match can be obtained.
{{{{{ Even if instead of taking the absorbing column density according
to equation (\ref{N1}), we assume an inclination angle $i= 42~^\circ$
and use equation (\ref{intNi}) we still get a good fit.
This result is illustrated by the dashed line in the lower right panel
of Figure \ref{flm}. }}}}}
In short, with the present model we already
($i$) strengthen previous results (e.g., Pittard et al. 1998;
Ishibashi et al. 1999; Corcoran et al. 2001a; Pittard \& Corcoran 2002)
that the X-ray emission can be explained by wind collision and
($ii$) achieve our main goal of explaining the
X-ray light curve, and in particular demonstrating that the shutting
down of the secondary wind might account for the X-ray minimum.

{{{ We refer now to the cycle-to-cycle variation (property 5 in section 2.1).
In the winds collision model the higher maximum X-ray luminosity prior to
the 2003.5 minimum compared to that prior to the 1997.9 minimum, can be accounted
for by either less obscuration by the dense primary wind in the 2003.5 minimum,
and/or a higher mass loss rate from the secondary in the 2003.5 minimum.
For example, if the primary has the same mass loss rate in the two minima,
but it is faster in that of 2003.5, then it has two effects to enhance
the observed X-ray flux: ($i$) it pushes the stagnation point closer to the
secondary where the secondary wind is denser, therefore radiates a higher fraction
of its thermal energy; and ($ii$) the higher velocity implies lower density,
thus less absorption.
The data are not of high enough quality to distinguish
between these two possibilities.
The qualitative similarity in the two minima is expected in our model,
as the binary parameters (stellar masses; eccentricity; orbital period)
did not change, and the differences in winds parameters are not sufficiently
large to affect the collapse of the winds interaction region.  }}}

It is not straight forward that our results of the X-ray emission
are similar to that of Corcoran et al. (2001) and Ishibashi et al.
(1999), because they use a formula (eq. 1 in Ishibashi et al.
1999) based on the results of Usov (1992). However, Usov (1992)
derived these formulae for binary systems with weaker winds. Near
periastron this equation (eq. 1 of Ishibashi et al. 1999) gives an
unabsorbed X-ray luminosity greater than the total kinetic power
in the two winds. This is the reason that we used a different
formula (our eq. \ref{lx2}).
We are gratified that we could account for the general behavior of the X-ray emission.

{{{ We thank an anonymous referee for very useful comments. }}}
This research was supported by the Israel Science Foundation,
grant 28/03 and by the Asher Space Research Institute.

\clearpage
\begin{figure}
\resizebox{0.66\textwidth}{!}{\includegraphics{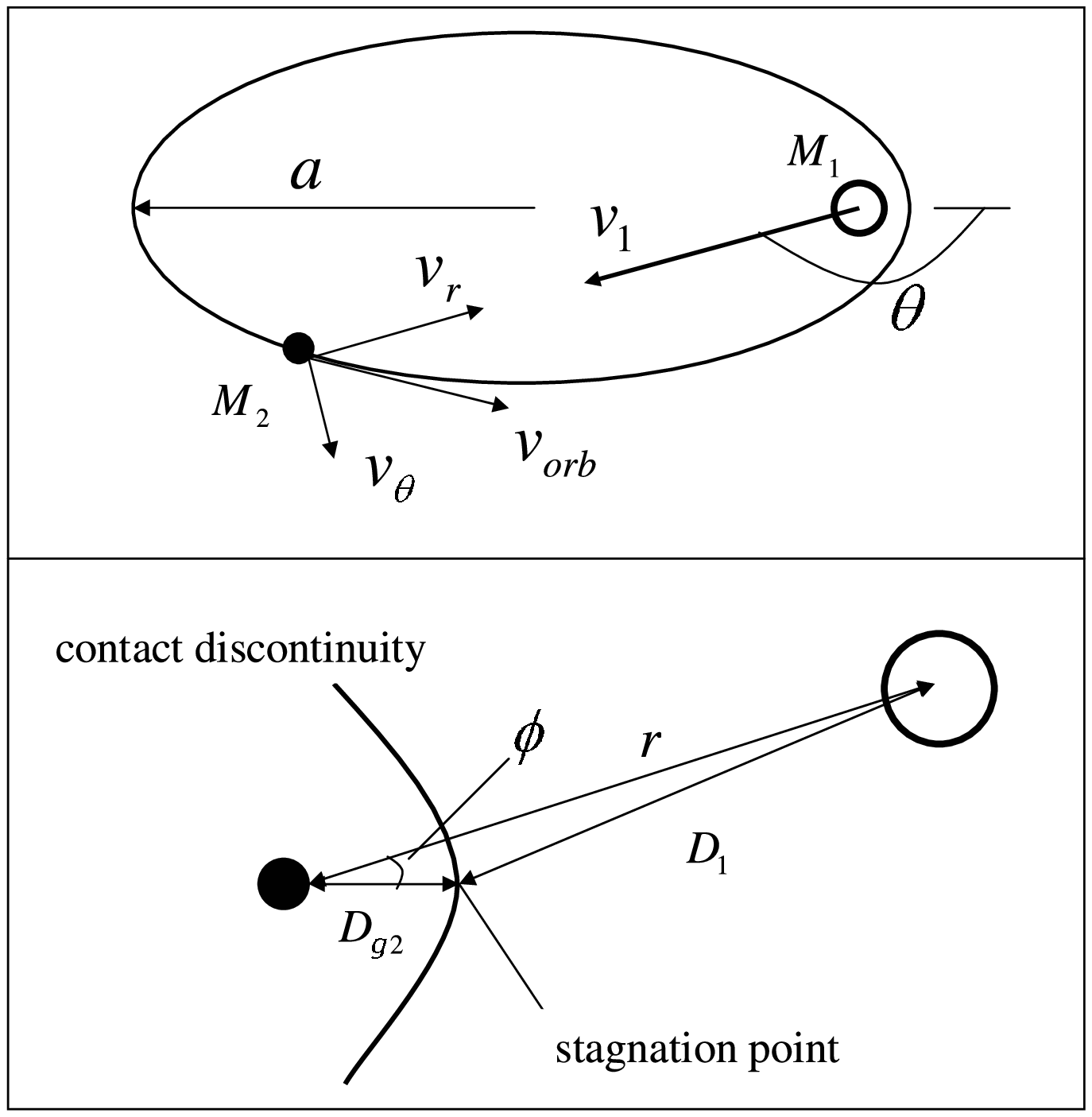}}
\caption{Upper panel: the orbit and relevant velocities in the
rest frame of the primary star of mass $M_1$. Drawn are the
primary's wind velocity $v_1$, the two stars relative orbital
velocity $v_{\rm orb}$, and $v_r$ and $v_\theta$ which are the
radial and tangential components of $v_{\rm orb}$, respectively.
Lower panel: Geometrical definitions relevant to the flow near the
stagnation point. The contact discontinuity is the surface where
the two winds meet after they have passed the shock waves. The
velocity directions in the lower panel are as in the upper panel
(the secondary moves to the lower right). From Soker (2005b). }
\label{forbit1}
\end{figure}
\begin{figure}
\resizebox{0.66\textwidth}{!}{\includegraphics{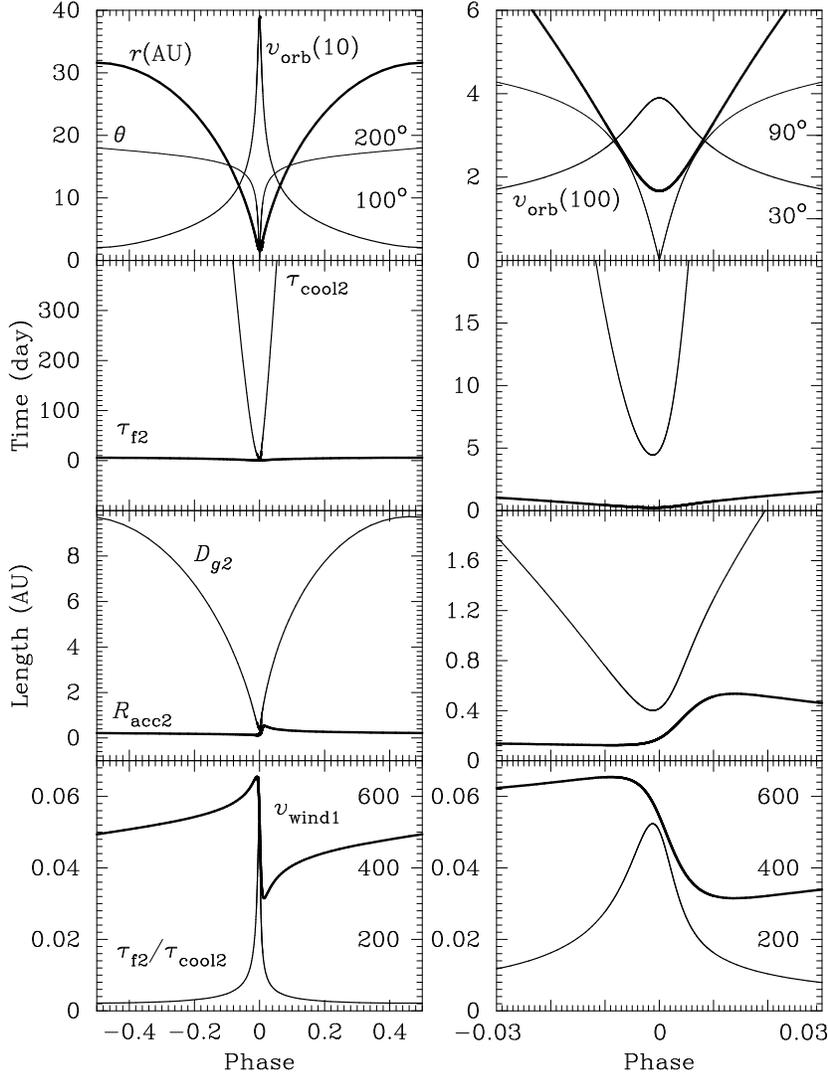}}
\caption{Several physical variables as function of orbital phase
(phase zero is at periastron). The left column covers the entire
orbit, while the column on the right covers the time just prior
and after periastron. Upper row: The orbital separation $r$ (in
AU), and the relative orbital speed of the two stars $v_{\rm orb}$
(in $10 \km \s^{-1}$ on the left and $100 \km \s^{-1}$ on the
right). The angle $\theta$ is the relative direction of the two
stars as measured from periastron (scale on the right in degrees).
Second row: The radiative cooling time of the shocked secondary's wind
near the stagnation point $\tau_{\rm cool2}$ (upper line), and the
typical time $\tau_{f2}$ for the shocked secondary wind to flow
out of the winds interaction zone (lower line).
Third row: the distance of the
stagnation point from the secondary when the gravity of the
secondary is included $D_{g2}$ (upper line; see Figure
\ref{forbit1}), and the Bondi-Hoyle accretion radius of the
secondary star $R_{\rm acc2}$ (lower line).
Lower row: The velocity of the primary wind relative to the
stagnation point $v_{\rm wind1}$ (thick line), and  the ratio
$\tau_{\rm f2}/\tau_{\rm cool 2}$ (thin line).
For more detail see Soker (2005b)}.
\label{forbit2}
\end{figure}
\begin{figure}
\resizebox{0.88\textwidth}{!}{\includegraphics{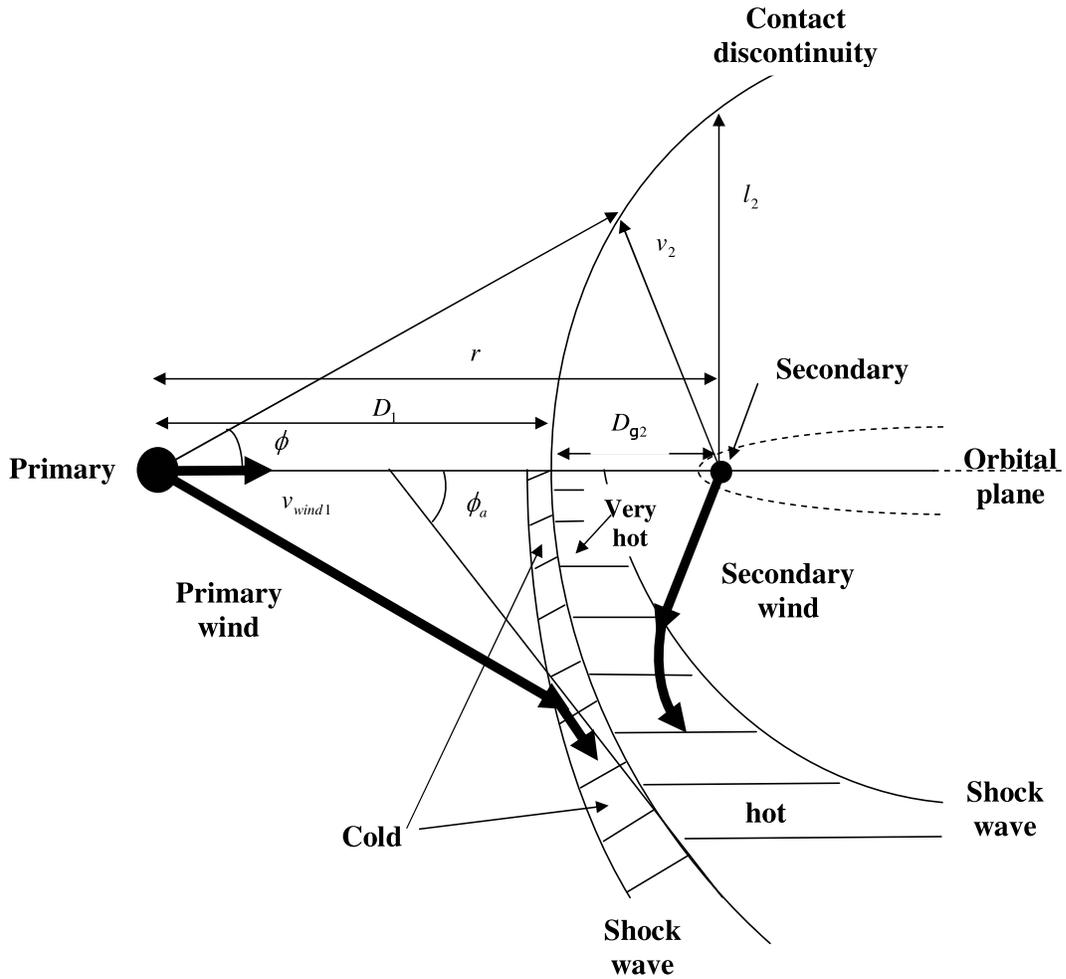}}
\caption{Schematic drawing of the collision region of two stellar
winds and definition of several quantities. The two thick lines
represent winds' stream lines. The two shock waves are drawn only
in the lower half. The post-shock regions of the two winds are
hatched.
The dashed line shows the accretion column which exists,
according to the proposed model, only for $\sim 70-80$~days
during the accretion period which corresponds to the X-ray  minimum.
This region is enlarged for this accretion period in figure \ref{facc}.
} \label{fwinds}
\end{figure}
\begin{figure}
\vskip -0.6 cm
\resizebox{0.68\textwidth}{!}{\includegraphics{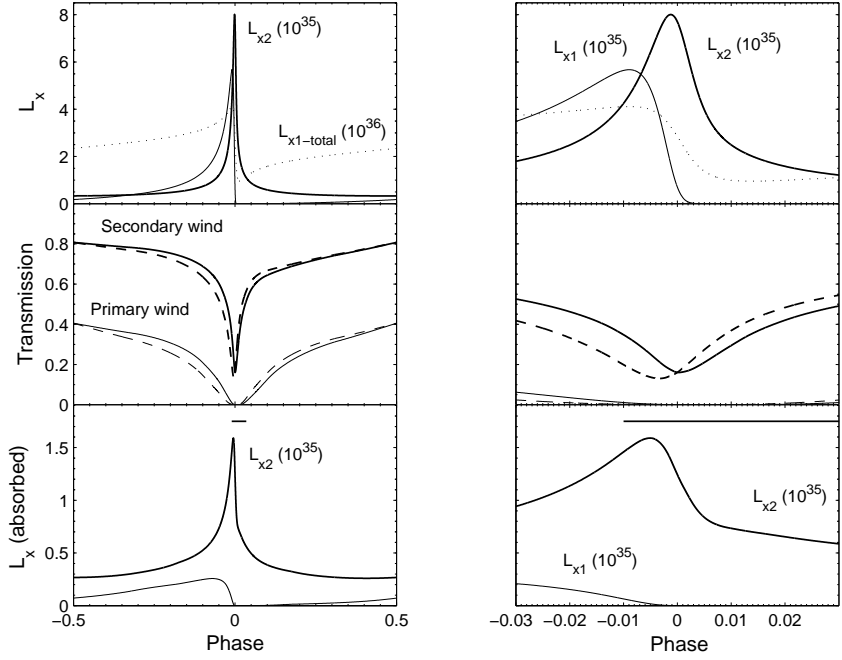}}
\vskip -0.2 cm
\caption{The X-ray luminosity outside the X-ray minimum. Upper
row: the dotted line shows the intrinsic (absorption not
included) luminosity of the shocked primary wind in the entire
$0.01-10 \kev$ band, in units of $10^{36} \erg \s^{-1}$.
The thin solid line shows the intrinsic
luminosity of the shocked primary wind only in the $2-10 \kev$
band $L_{x1}$ (eq. \ref{lx1}), and the thick line shows the
intrinsic luminosity of the shocked secondary wind only in the
$2-10 \kev$ band $L_{x2}$ (eq. \ref{lx22} with $k_2=2$), both in
units of $10^{35} \erg \s^{-1}$.
{{{{{ Most of the contribution to $L_{x1}$ is in the $2-3 \kev$ band,
explaining its large attenuation.
Middle row: The X-ray transmission factor through the primary wind
for the $2-10 \kev$  band, calculated with XSPEC.
Solid lines show the transmission using the column density calculated by
equation (\ref{intN1}) (or eq. \ref{N1}),
while the dashed lines show the transmission for an inclination
$i=42^\circ$ and periastron orientation $\omega=180^\circ$ (equation \ref{intNi}));
$k_g=0.5$ in all cases.
The two upper lines show the transmission of the X-ray emitted by the shocked
secondary wind, while the two lower lines show the transmission of the
X-ray emitted by the shocked primary wind.
These lines emphasize the small difference between the two absorbing
gas geometries considered here, and the high absorption of the
X-ray emitted by the shocked primary wind. }}}}}
Lower row: The calculated X-ray luminosities of the shocked two winds in
the $2-10 \kev$ band when attenuation by equation (\ref{intN1}) is
included; thin and thick lines, respectively, represent the
attenuated X-ray luminosities of the shocked primary and secondary
winds. These quantities should be compared with observations
outside the X-ray minimum.
{{{{{ The horizontal line in each of the two panels in the lower row
marks the time during which according to our model the secondary wind
does not exist, or is highly suppressed.
It spans the time starting a little before the minimum and ending a little after
the minimum, phase $-0.01$ to $0.035$. }}}}}}
\label{flumx}
\end{figure}
\begin{figure}
\resizebox{0.68\textwidth}{!}{\includegraphics{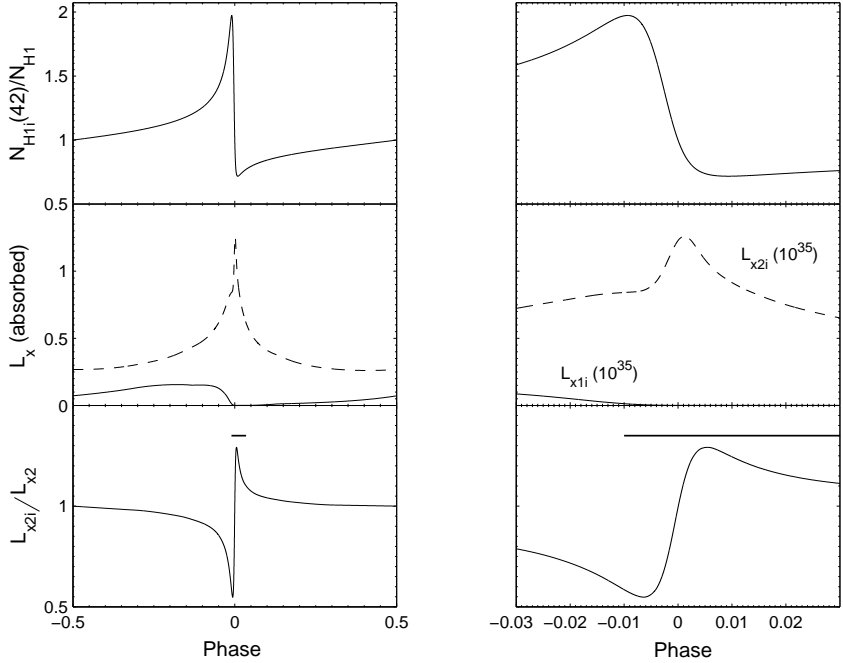}}
\caption{ {{{{{ Upper row: the ratio $N_{H1i}(i=42^\circ)/N_{H1}$, namely
of the column densities in the two geometries considered here,
where $N_{H1i}(i=42^\circ)$ is from equation (\ref{intNi}) and
$N_{H1}$ is from equation (\ref{intN1}), or scaled in equation (\ref{N1}).
Middle row: Like the lower row in Figure \ref{flumx} but the observed X-ray luminosities
$L_{x2i}$ and $L_{x1i}$ are calculated with attenuation by the column density
$N_{H1i}(i=42^\circ)$ with $k_g=0.5$.
Lower row: the ratio $L_{x2i}/L_{x2}$. This ratio shows that the differences
between the two geometries considered here are small.
The horizontal lines in the last row have the same meaning as in Figure \ref{flumx} }}}}}  }
\label{fincline}
\end{figure}
\begin{figure}
\resizebox{0.77\textwidth}{!}{\includegraphics{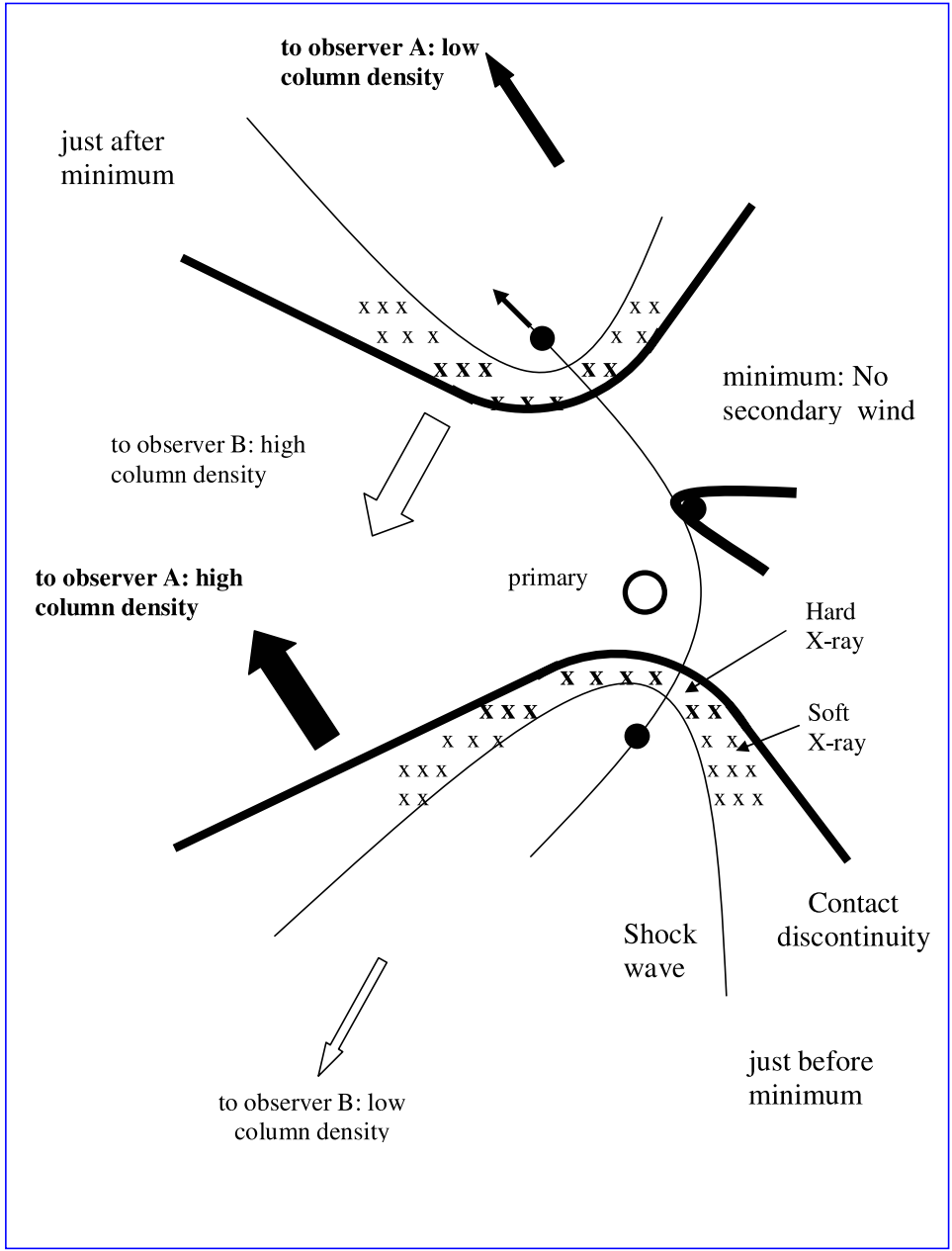}}
\caption{ Schematic drawing (not for scaling and not the exact
shock waves and contact discontinuity structures) of the flow
structure at three epochs: Just before and after the X-ray
minimum when the two wind exist, and during the X-ray minimum,
when the secondary wind is assumed to be extinct. Note that
according to our model the X-ray minimum is not symmetric about
periastron. The shocked primary wind is marked by the thick arcs;
the X-ray emitting shocked secondary wind is in the region marked
by `x's'; the open and filled circles mark the positions of the
primary and secondary, respectively. Two possibilities for the
orientation of the system are represented by the arrows (see
text). } \label{fabsorb}
\end{figure}
\begin{figure}
\resizebox{0.88\textwidth}{!}{\includegraphics{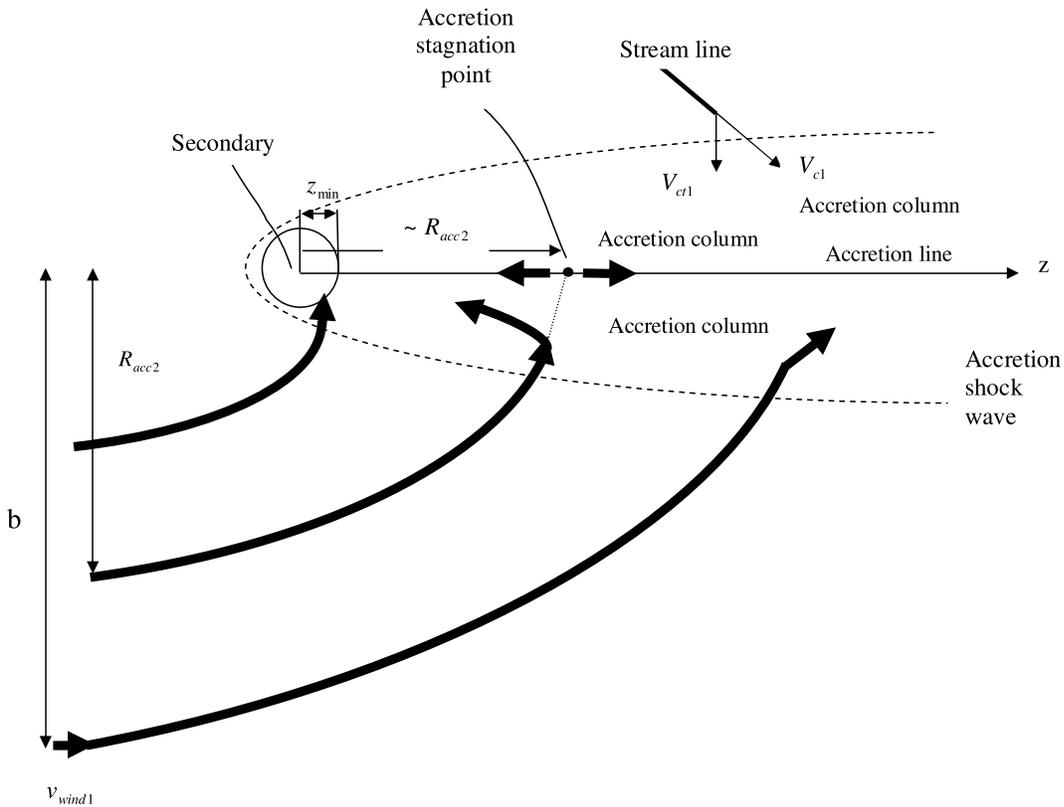}}
\caption{Schematic drawing of the isothermal high-Mach number
Bondi-Hoyle-Lyttleton type accretion flow, and the definition of
some quantities used in the text. We suggest that this type of
accretion occurs during the X-ray minimum, and we further assume
it shuts down the secondary wind. The stagnation point along the
accretion line should not be confused with the stagnation point of
the colliding winds. } \label{facc}
\end{figure}
\begin{figure}
\resizebox{0.88\textwidth}{!}{\includegraphics{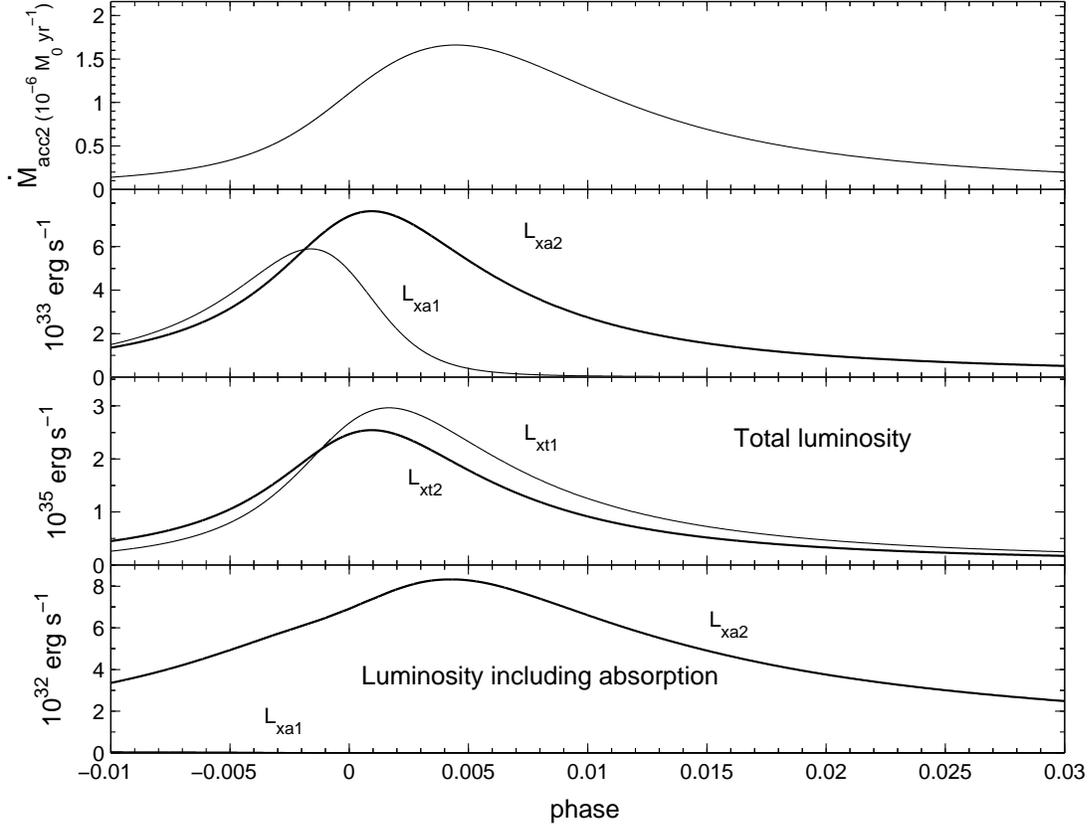}}
\caption{Quantities related to the X-rays emitted by primary wind
material accreted onto the secondary during the X-ray minimum.
Upper panel: The accretion rate $\dot M_{\rm acc2}$ (eq.
\ref{macc2}). Second panel: The thick line shows the X-ray
luminosity in the $2-10 \kev$ band of material accreted by the
secondary through the accretion column $L_{xa1}$ (eq. \ref{lxa1}).
The thin line shows the X-ray luminosity in the $2-10 \kev$ band
of material accreted directly by the secondary $L_{xa2}$
(\ref{lxa2s}). Third panel: The total luminosity of the shocked
accreted gas, namely, taking $F_{210}=1$ in equations (\ref{lxa1})
and (\ref{lxa2s}). Lower panel: The luminosities shown in the
second panel but with absorption included.
{{{{{ Note that the primary wind X-ray emission is completely absorbed. }}}}} }
\label{flacc}
\end{figure}
\begin{figure}
\resizebox{0.88\textwidth}{!}{\includegraphics{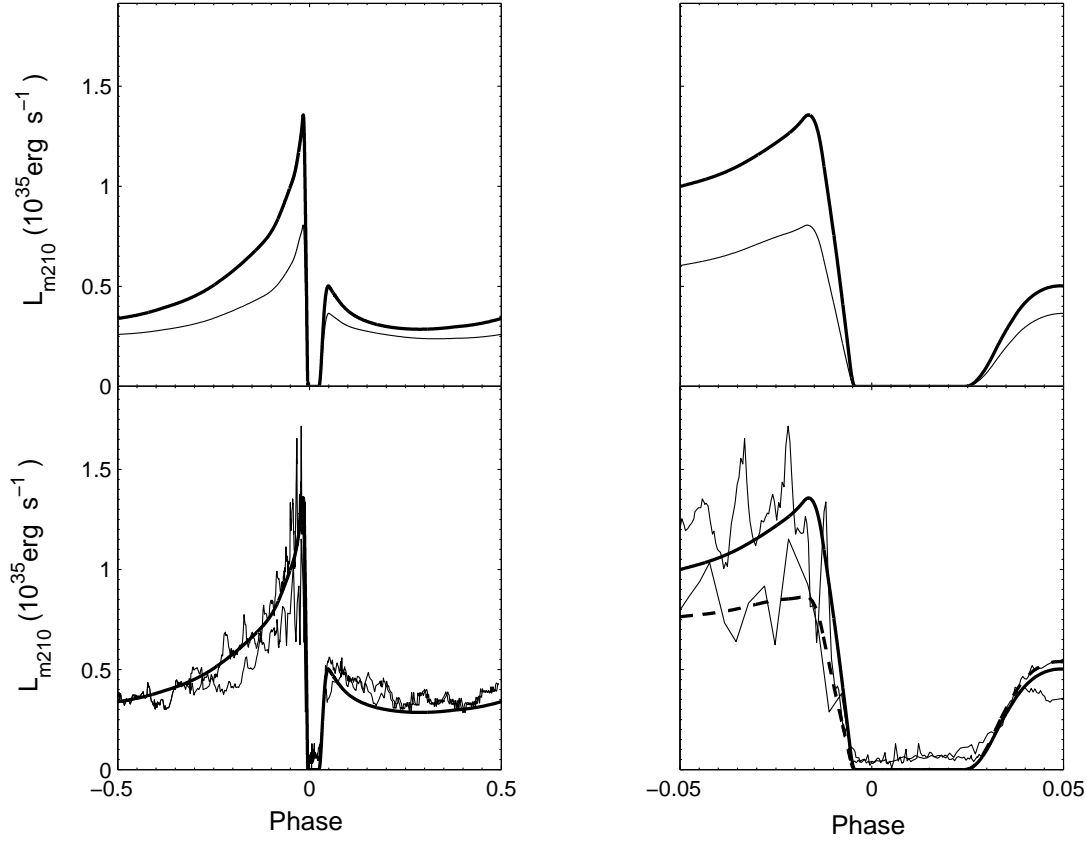}}
\caption{Upper row:
The absorbed X-ray luminosity between $2-10 \kev$ according to our model,
namely, equation (\ref{lm}), but with absorption included
and not considering the residual emission $L_{x-{\rm res}}$ .
A value of $k_2=2$ is assumed in equation (\ref{lx2}).
The absorbing column density is according to equation (\ref{N1}),
with $k_g=0.5$ (thick line) and $k_g=1$ (thin line).
Lower row:
The model absorbed X-ray luminosity with $k_g=0.5$ (thick line),
and the observed X-ray luminosity from Corcoran (2005) in the
$2-10 \kev$ band for a period covering two X-ray minima (two thin lines).
{{{{{ The dashed line in the right lower panel shows the X-ray emission
for $\omega=180^{\circ}$ and $i=42^{\circ}$ (see eq. \ref{intNi}). }}}}}}
\label{flm}
\end{figure}
%
\end{document}